\documentclass[epjST]{svjour}
\usepackage{graphics}
\usepackage{amsmath,amssymb,dsfont}
\newcommand{\be}{\begin{equation}}
\newcommand{\ee}{\end{equation}}
\newcommand{\bea}{\begin{eqnarray}}
\newcommand{\eea}{\end{eqnarray}}
\newcommand{\lbl}{\label}

\newcommand{\nn}{\nonumber}
\newcommand{\re}{\mathrm{e}}

\newcommand{\borel}{\mathcal{B}^{[\rho]}(\sigma)}

\newcommand{\bl}{\mathcal{B}_{<}^{[\rho]}(\sigma)}
\newcommand{\bh}{\mathcal{B}_{\mathrm{>}}^{[\rho]}(\sigma)}

\newcommand{\s}{\sigma}
\newcommand{\sh}{\widehat{\sigma}}

\begin{document}
\title{Violation of Quark-Hadron Duality}
\subtitle{The missing oscillation in the OPE}
\author{Santiago Peris \inst{1,}\inst{2}\fnmsep\thanks{\email{peris@ifae.es}} }
\institute{Grup de Fisica Teorica, Dept. de Fisica\\
Univ. Autonoma de Barcelona, \\
08193 Bellaterra (Barcelona) Spain
\and Institut de Fisica d’Altes Energies (IFAE) and  \\
The Barcelona Institute of Science and Technology,\\
Campus UAB, 08193 Bellaterra (Barcelona) Spain
 }
\abstract{
The origin of quark-hadron Duality Violations (DVs) can be related to the sigularities of the Laplace transform of the spectral function. With the help of rather generic properties of the large-$N_c$ approximation and a generalized form for the radial trajectories found in Regge Theory, we may locate these singularities in the complex plane and obtain an expression for the DVs which turns out to agree with general expectations. Using the two-point vector correlator as a test laboratory, we show how the usual dispersion relation may give rise to perturbation theory, the power corrections from the condensate expansion and DVs.
} 
\maketitle
\section{Introduction}
\label{intro}
Quark-hadron Duality Violations (DVs) refer to the inability of the condensate expansion to reproduce the spectral function, even at relatively large energies.\footnote{In what follows we will consider the perturbative expansion as the OPE contribution from the unit operator.} This is in contradistinction to the Euclidean regime, where the OPE is found to give a rather accurate description of the QCD correlators even down to scales of $\sim 1\, \mathrm{GeV}$ \cite{PPR}.

Right from the start, it is fair to say that there is at present no first-principle understanding of quark-hadron Duality Violations (DVs) from QCD \cite{Shif}-\cite{Shif1}. Since the pioneering work of Poggio, Quinn and Weinberg \cite{PQW}, our understanding of DVs has been based on a set of increasingly sophisticated models which emphasize those features of DVs which were considered relevant to the problem at hand, however at the expense of missing some other known properties of QCD. For instance, one could obtain a nonzero contribution from DVs, but with the wrong perturbative series in $\alpha_s$ \cite{Shifman}.

The present text is based on the recent analysis done in Ref.\,\cite{us}.  This analysis can be considered a step forward in the direction of bridging the gap between the accumulated knowledge from these models and QCD in that it reconciles  models of DVs with several relevant ingredients such as, for instance, perturbation theory. In particular, as we shall see, DVs are found \emph{together} with a perturbative theory capable of having the expected renormalon structure. Our approach will be heavily based on the picture of QCD given by the $1/N_c$ expansion (and its expected Regge properties),  as this interplay between perturbative and nonperturbative aspects of QCD clearly requires an approach which goes beyond the perturbative framework. Furthermore, recent developments in the Theory of Transseries and Hyperasymptotics will also be very useful to set the right mathematical language \cite{math}.

Let us begin by considering the scalar function $\Pi(q^2)$ of the vector-current two-point  correlator in QCD as a test laboratory. An important property of this function is that it is analytic in the complex $q^2$ variable except for a cut for $q^2>0$ where the spectral function lies.\footnote{We will simplify our world by taking the chiral limit.}  To get rid of a scheme dependent additive constant, it is convenient to consider the Adler function, defined as
\be\label{eq:A}
\mathcal{A}(q^2)=-q^2\frac{d}{dq^2}\Pi(q^2)\ .
\ee
Asymptotic freedom guarantees that this function can be calculated with the OPE in terms of quarks and gluons provided the relevant momentum in the process, $q^2$, is large. However, the way this regime is attained is very different whether we are looking at a positive or negative $q^2$, even though, one might naively think that these two $q^2$-regimes should be connected by analytic continuation as follows from the analyticity of $\Pi(q^2)$. For instance, taking $q^2<0$ (Euclidean) and large, we can approximate $\mathcal{A}(q^2)$
as\footnote{For simplicity, we are normalizing the parton model contribution to unity.}
\be
\lbl{eq:QCD}
\mathcal{A}(q^2)_{\rm OPE}=1+ \sum_{n\ge 1} a_n\, \alpha^n_s(-q^2)
+\sum_{n\ge 1} \frac{b_n(q^2)}{(-q^2)^n}\ ,
\ee
where the first and second terms correspond to the perturbative
series in powers of the running strong coupling $\alpha_s(q^2)$
and the third term may be associated with the nonperturbative condensate expansion
of the OPE. The coefficients $b_n(q^2)$ depend logarithmically on $q^2$ but this dependence is screened by at least one extra power of $\alpha_s$ and it is, therefeore, a reasonable simplification to neglect it. In this case, using the known prescription to pick the imaginary part as $\mathrm{Im}(q^2+ i \epsilon)^{-1}=-i \pi \delta(q^2)$, and derivatives thereof,  one discovers that the whole  imaginary part in Eq.\,(\ref{eq:QCD}) becomes purely perturbative and all the nonperturbative power corrections vanish.\footnote{Even if we consider the log corrections in the $b_n$ coefficients, at high enough $q^2$, a finite number of these power corrections cannot give rise to the oscillations commonly seen in the spectral data. } In other words, the OPE does not exhibit the analyticity of $\Pi(q^2)$. Phenomenologically this means that the OPE misses all the resonance structure in the spectrum. The missed,
resonance-induced corrections become negligible only at very high energies, where the resonances become very broad and overlapping. This is the origin of DVs.

In order to understand where these DVs come from, it is clear that we need a framework in which the analytical properties of the Adler function are manifest. Therefore we will start with the dispersion relation
\bea
\lbl{eq:dispersion}
\mathcal{A}(q^2)&=&q^2 \int_0^\infty dt\ \frac{\rho(t)}{(t-q^2-i\epsilon)^2}\nn \\
&=& -\, q^2 \int_0^{\infty} d\sigma\ \mathrm{e}^{\sigma q^2}
\, \sigma \borel\
\eea
where $\rho(t)=\frac{1}{\pi}\mathrm{Im}\Pi(t)$ and
\be
\lbl{eq:borel}
\borel=\int_0^{\infty}dt\  \rho(t)\  \mathrm{e}^{-\sigma t}\ .
\ee
The analytical properties of $\mathcal{A}(q^2)$ in the $q^2$ complex plane follow directly from
those of ${\cal{B}}^{[\rho ]}(\sigma )$ in the complex $\sigma$ plane, a property which will become crucial in what follows.

Let us point out a few (simple but)  important properties expressed in Eqs. (\ref{eq:dispersion}) and (\ref{eq:borel}). First, this representation is valid for $\mathrm{Re}\, q^2<0$ and is exact. Furthermore, the Adler function is dimensionless so this means that the actual dependence is on $q^2/\Lambda_{QCD}^2$ where $\Lambda_{QCD}$ is the dynamical mass scale of QCD. In other words, Eq. (\ref{eq:dispersion}) really makes sense only after  $\Lambda_{QCD}$ (or a convenient proxy for it) has been identified.  There is no trace of $\alpha_s$ in these equations, so this is an auxiliary parameter whose job is to encapsulate some of the $q^2$ dependence (i.e., the logs). The function $\borel$ is  analytic in the complex $\sigma$ plane for $\mathrm{Re}\,\sigma > 0$  (we may normalize  $\rho(t)\to 1$ as $t\to \infty$). Singularities of $\borel$, which will soon become crucial in our discussion, necessarily have to lie in the $\mathrm{Re}\,\sigma \leq 0$ half plane. The OPE is obtained by expanding $\borel$ about $\sigma=0$, where powers of $\sigma$ correspond to inverse powers of $-q^2$ and logarithms of $\sigma$ corresponds to logarithms of $-q^2$. In particular $\sigma\borel \to 1$ as $\sigma \to 0$,  if $ \mathcal{A}(q^2)\to 1$ as $-q^2\to \infty$. It is because $\rho(t)$ extends all the way to infinity that the OPE does not define an analytic function. If the spectrum were to vanish starting from some finite $t_0$ onwards, the large $q^2$ expansion would have a disk of convergence for $|q^2| > t_0$. Consequently, it is expected that the OPE is (at most) an asymptotic expansion and, as such, it will miss terms $\mathcal{O}\left(e^{-|q^2|/\Lambda_{QCD}^2}\right)$, just as the asymptotic perturbative expansion misses terms  $\mathcal{O}\left(e^{-1/\alpha_s}\right)$ due to  renormalons. DVs are expected to exhibit this exponential suppression, modulated by some sort of oscillation as seen in the data. Finally, notice that Eq. (\ref{eq:dispersion})  is equally valid for $\sigma >0$ and $q^2<0$, as for $\sigma<0$ and $q^2>0$.

It is this last observation that says that, if we can analytically continue $\borel$ to the $\mathrm{Re}\, \sigma <0$ from the $\mathrm{Re}\, \sigma >0$ half plane,  we could use Eq.\,(\ref{eq:dispersion}) to analytically continue the OPE from the Euclidean regime $q^2\to -\infty$ to the Minkowski regime $q^2\to \infty$ \cite{math}. By considering different rays in the $\sigma$ complex plane starting at the origin, instead of just the real axis as in Eq. (\ref{eq:dispersion}), we can analytically continue the Adler function in the $q^2$ complex plane provided $\frac{\pi}{2}< \arg \sigma + \arg q^2 <\frac{3\pi}{2}$ so that $e^{\sigma q^2}$ conveniently falls off at infinity. By rotating the $\sigma$ ray clockwise we rotate the $q^2$ region anti-clockwise (and viceversa).\footnote{In fact, to fulfill the reality property $\mathcal{A}({q^2}^*)=\mathcal{A}(q^2)^*$, one should rotate $\sigma$ clockwise for $\mathrm{Im}\, q^2>0$ and anti-clockwise for  $\mathrm{Im}\, q^2<0$.}

According to Cauchy's theorem, any two such $\sigma$ rays, differing by a finite angle, will yield the same result for the OPE of $\mathcal{A}(q^2)$ provided they do not enclose any singularities of $\borel.$\footnote{We will assume that $e^{\sigma q^2}\sigma\, \borel$ always falls off for to zero for $|q^2|\to \infty$.  } On the other hand if, in the process of this rotation in the $\sigma$ plane, we encounter singularities of $\borel$ the two OPE expansions for $q^2>0$ and $q^2<0$ will differ. This difference is nothing other than the DVs. The origin of DVs, therefore, lies in the location of the singularities of the $\borel$ function. In the following we will argue that these singularities are expected to be along the $\sigma$ imaginary axis at $N_c=\infty$ and slightly to the left of the imaginary axis when $N_c$ is large but finite.\footnote{In this context, $N_c=3$ is large enough.}

\section{Approaching $\mathbf{N_c=\infty}$ QCD}
\label{sec:1}
\subsection{Warm-up exercise: a simplified version of $\mathbf{N_c=\infty}$ QCD }
\label{sec:2}

\begin{figure}
\begin{center}

\resizebox{.8\columnwidth}{!}{%
\includegraphics{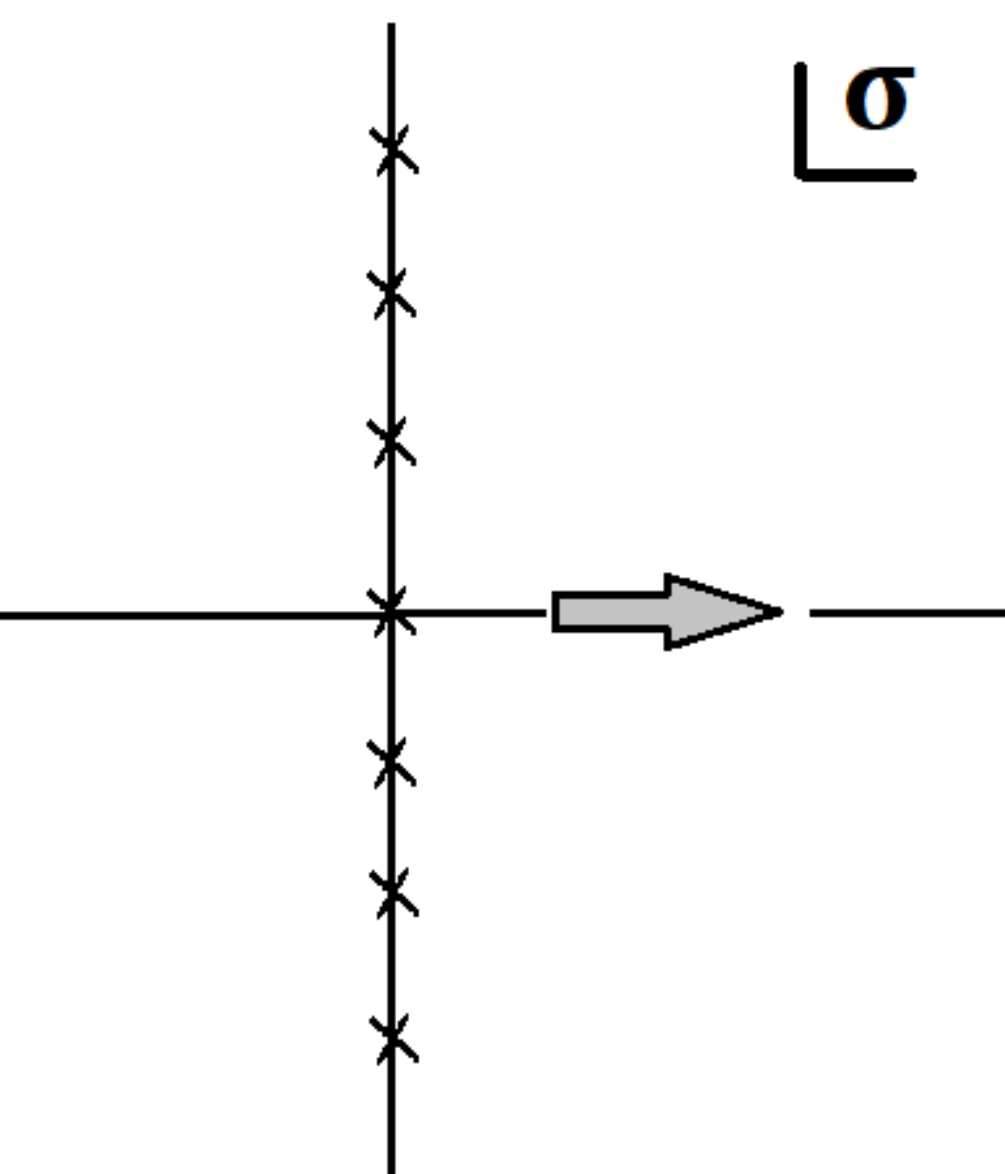}
\hspace{5in}
\includegraphics{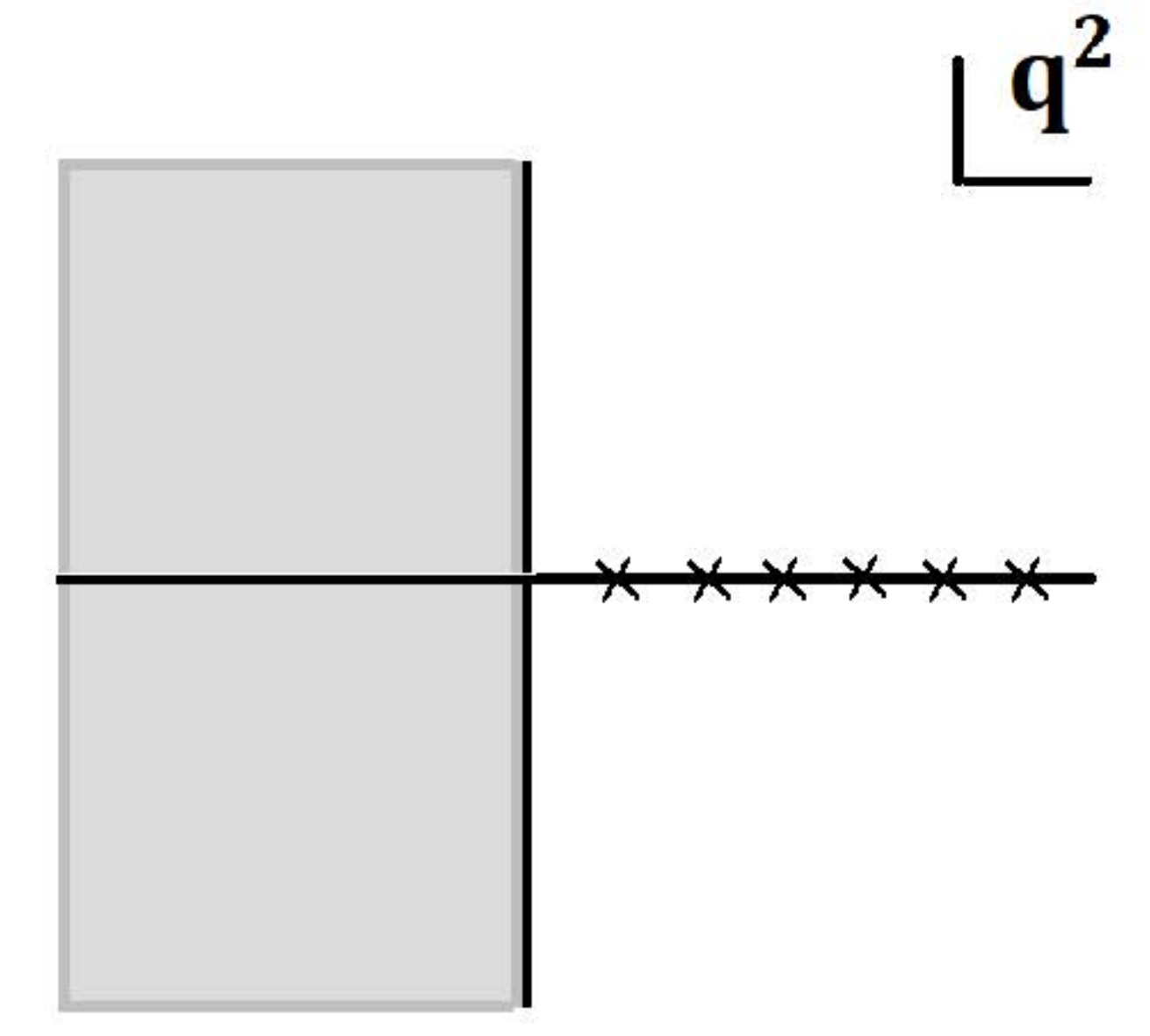}}

\resizebox{0.8\columnwidth}{!}{%
\includegraphics{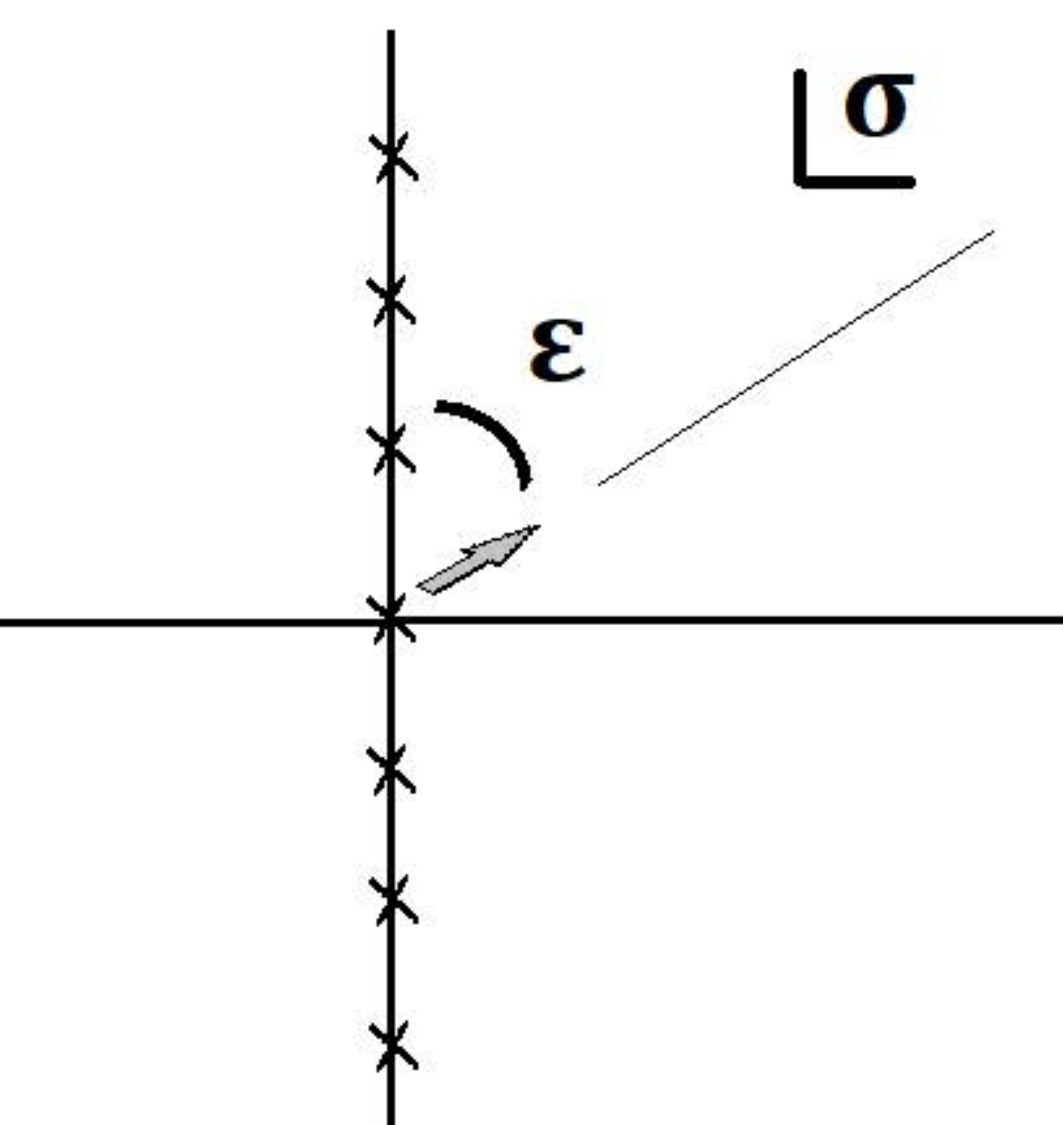}
\hspace{5in}
\includegraphics{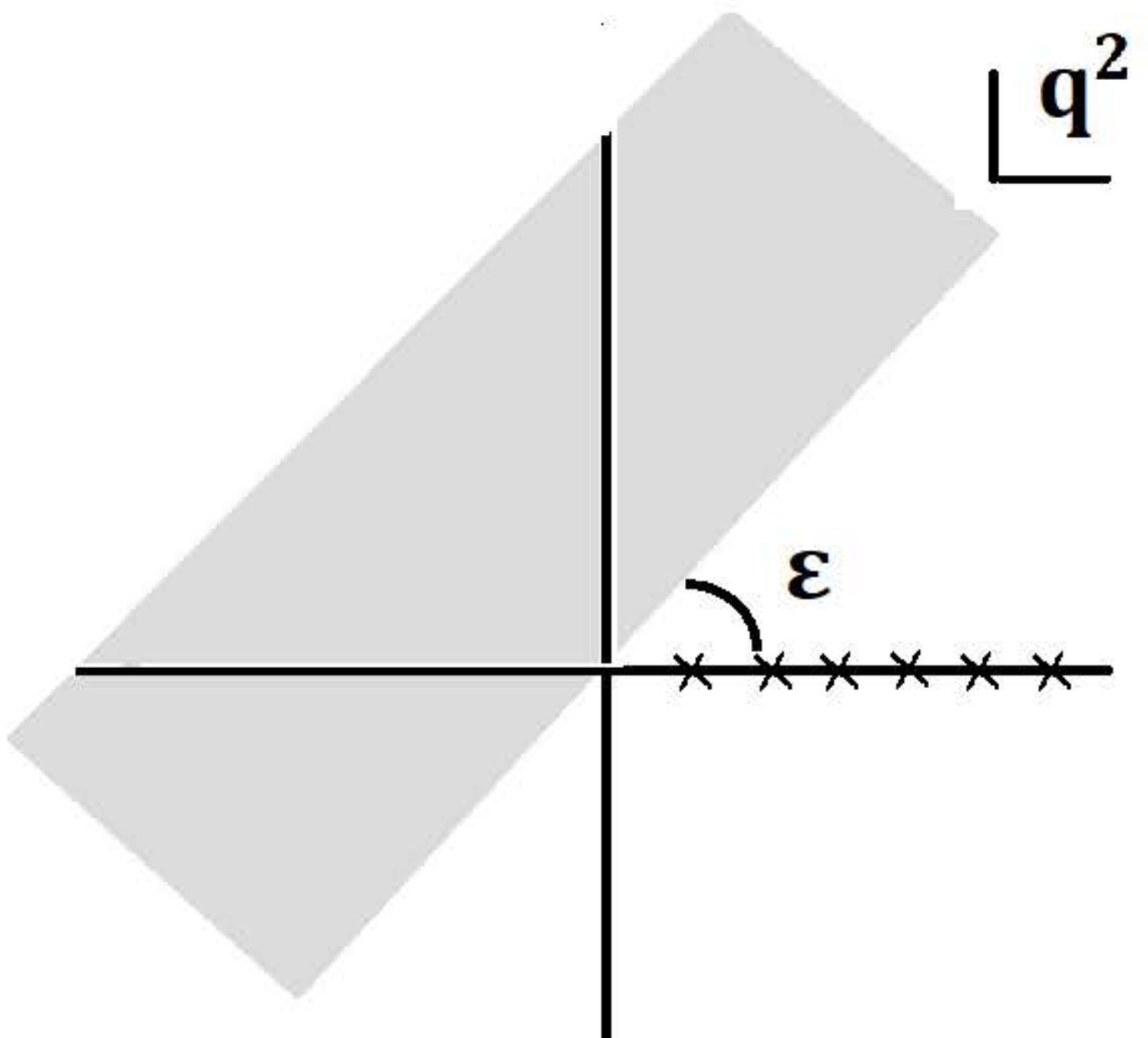}}

\resizebox{0.8\columnwidth}{!}{%
\includegraphics{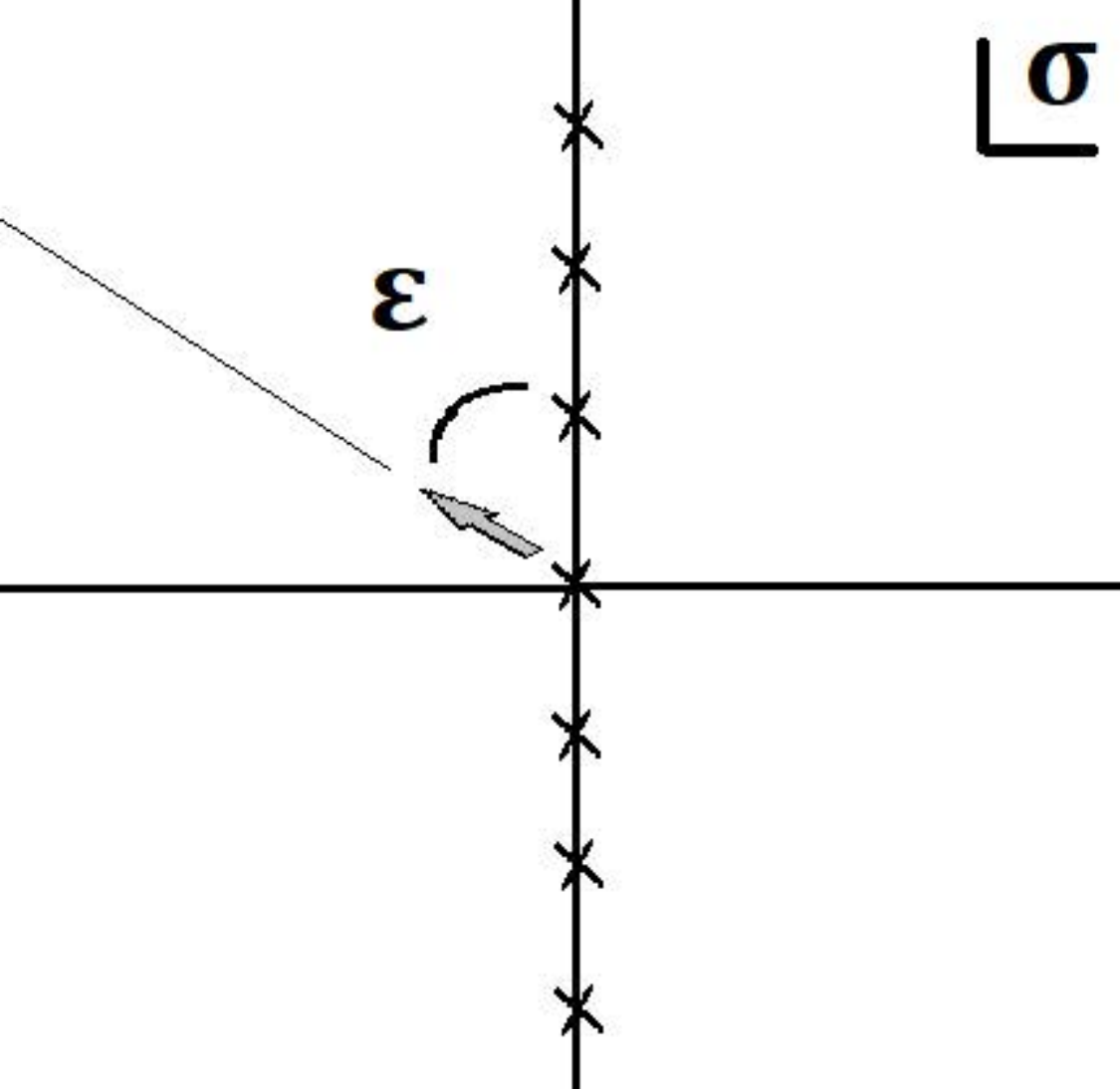}
\hspace{5in}
\includegraphics{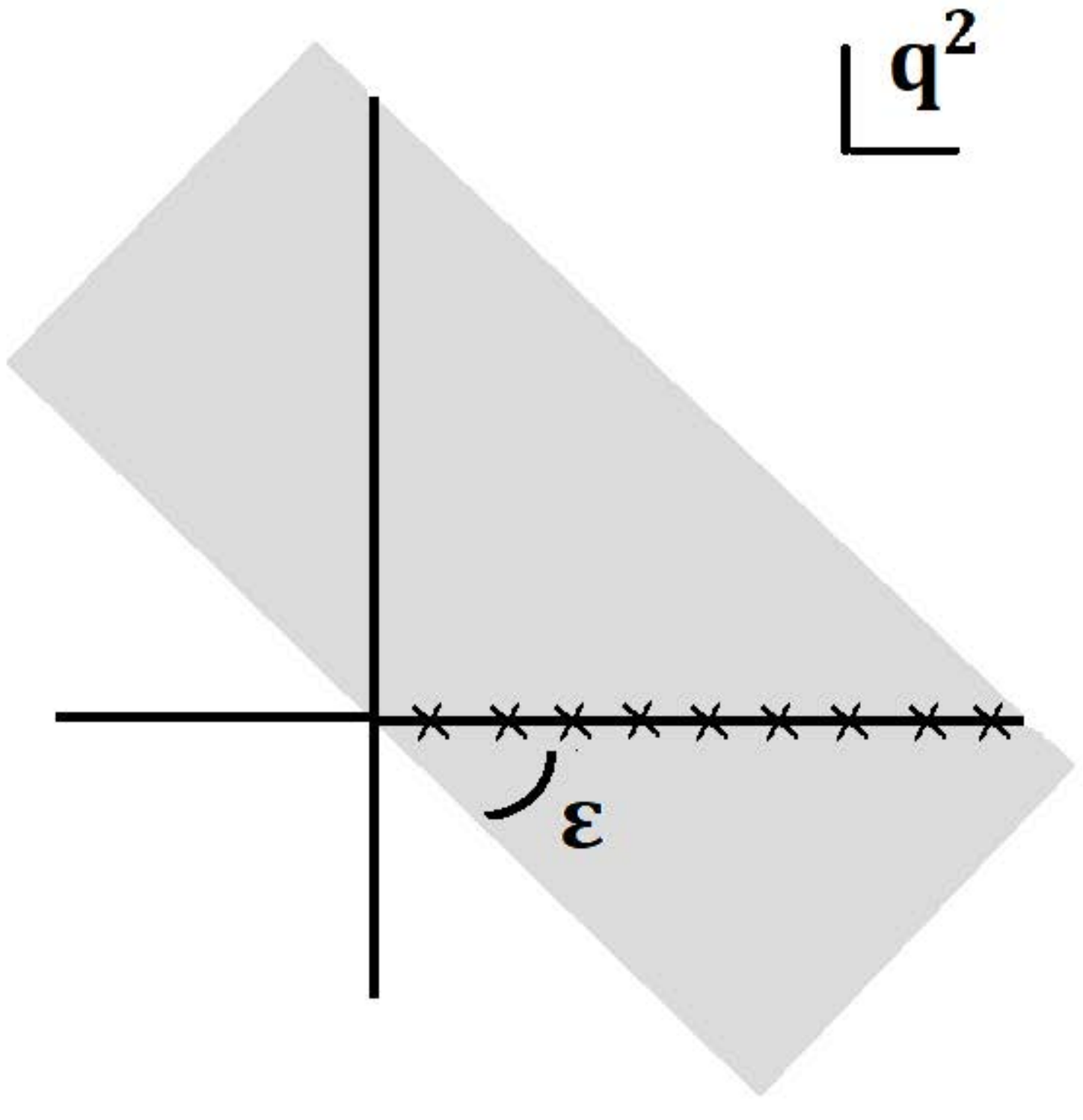}}

\resizebox{0.8\columnwidth}{!}{%
\includegraphics{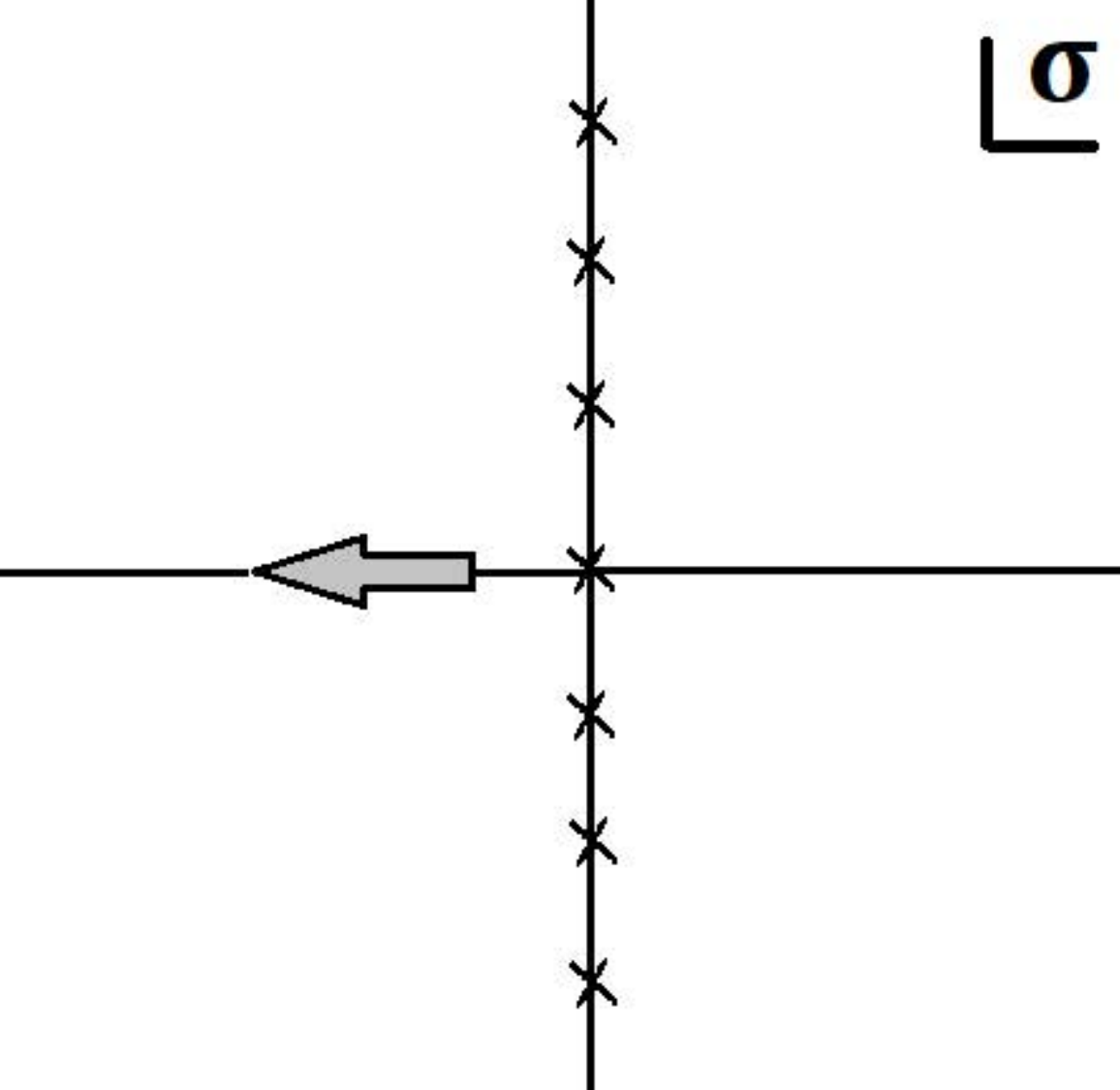}
\hspace{5in}
\includegraphics{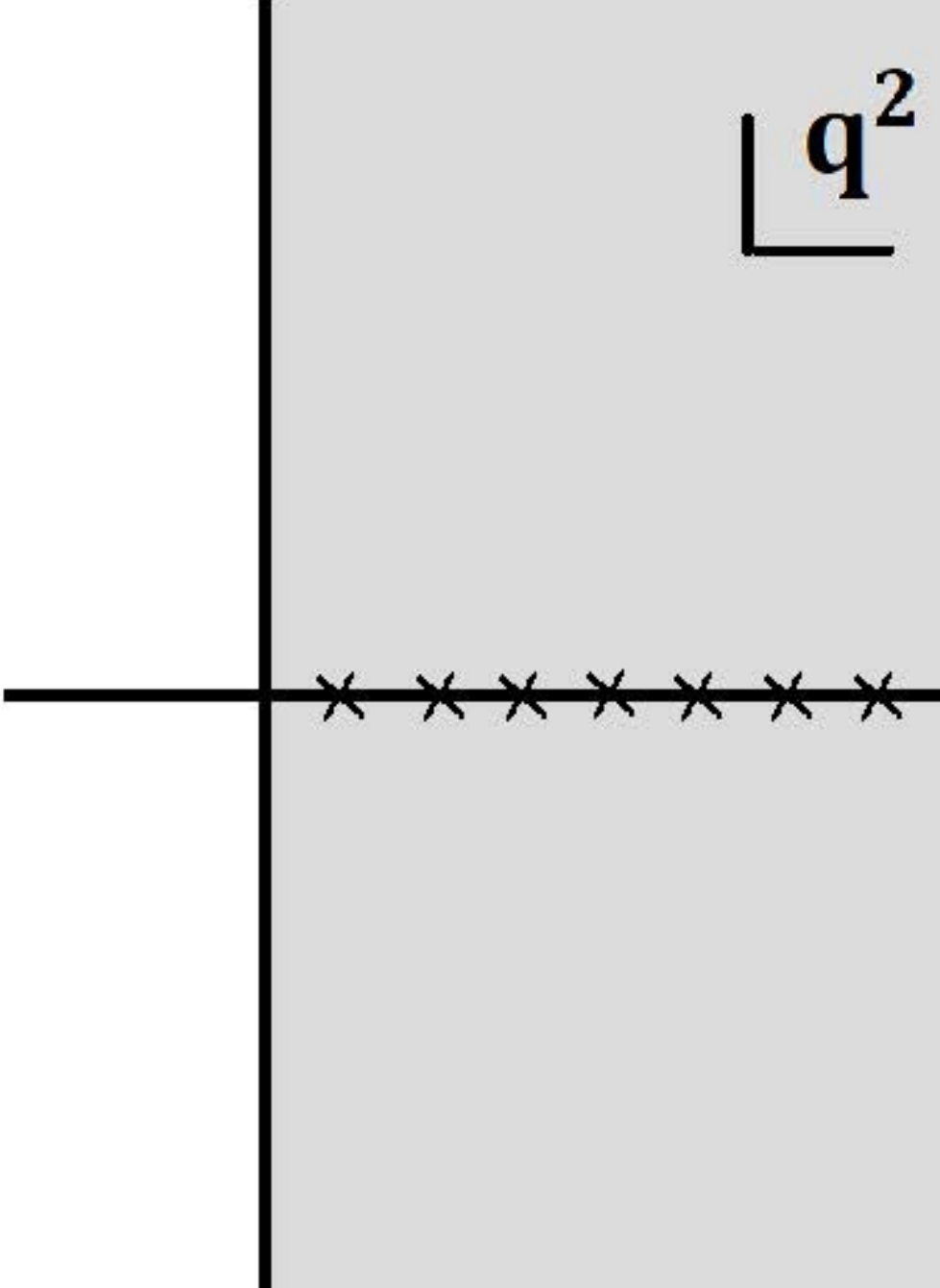}}

\resizebox{0.4\columnwidth}{!}{%
\includegraphics{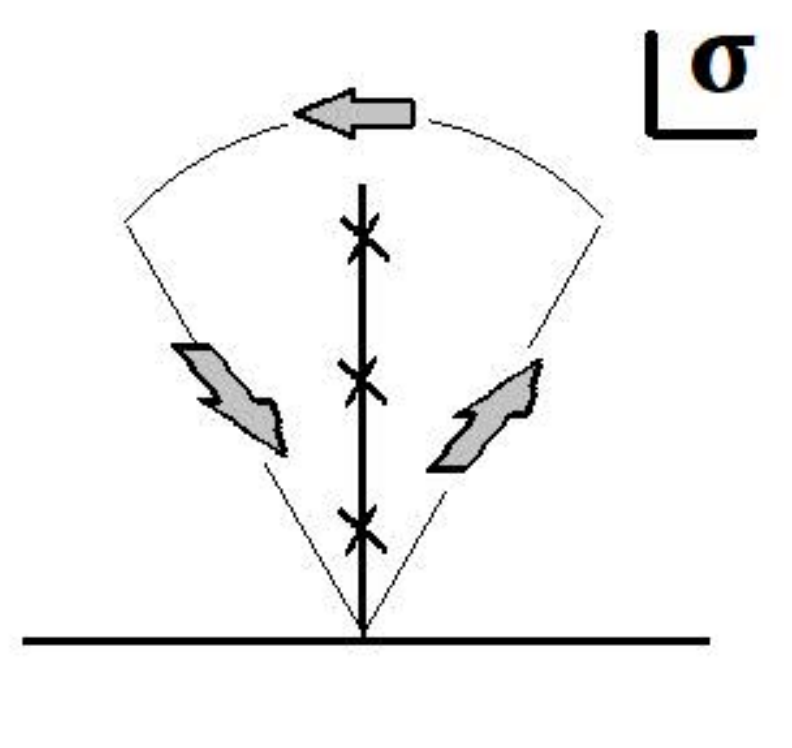}}
\end{center}
\caption{Analytic continuation using the
transform in Eq.\,(\ref{eq:dispersion}) and the  $\borel$ function, as discussed in the text,  in the large-$N_c$ limit.
Panels show, horizontally,  the $\sigma$ ray and the associated region covered in $q^2$ while, from top to bottom, we show the result of rotating the $\sigma$ ray anticlockwise. Crosses stand for poles (or branch points) of the $\borel$ function and the associated singularities of the Adler function in the $q^2$ plane. The bottom panel shows the contribution from the singularities as the $\sigma$ ray is rotated.}
\label{fig:1}
\end{figure}

QCD in two dimensions shows asymptotic linear Regge behavior in the spectrum at $N_c=\infty$ \cite{tHooft2d,CCG} and this property is believed to hold true also in four dimensions. Furthermore, this idea is supported by the string picture \cite{string} and also phenomenology \cite{Masjuan}. Therefore, it seems reasonable to start with a  simple spectrum given by
\be
\lbl{rhotoy}
\rho(t)=\sum_{n=1}^{\infty} F(n)\delta\!\left(t-M^2(n)\right)\quad,
\quad  n=1,2,3,\dots\ ,
\ee
with
\be
\lbl{massdecay}
M^2(n)=\Lambda^2 n\  , \qquad F(n)=F^2\ ,
\ee
in units where  $\Lambda=1$, and the spectral function is normalized to $F=1$.\footnote{We will use these units for the rest of this article.} Later on we will explore the consequences of relaxing these simplifications. Using this spectrum, the function $\sigma \borel$ is immediately calculated as
\bea
\lbl{rhohattoy}
\sigma \borel= \frac{\sigma}{\re^{\sigma}-1}
=\sum_{n=0}^\infty \frac{B(n)}{n!}\,\sigma^n
 \  ,
\eea
with $B(0)=1, B(1)=-1/2$ and
$B(2n> 1)=\frac{(-1)^{n+1} 2 (2n)!}{(2\pi)^{2n}}\, \zeta(2n)$, which are the  Bernoulli numbers and $\zeta(s)$ is the Riemann $\zeta$-function. As one can see, the function in Eq.~(\ref{rhohattoy}) has simple poles at $\sigma=\pm  2k\pi i\  (k=1,2,3,\dots)$, and residues $\pm  2k\pi i$. They are shown as crosses in the left panels
of Fig.\,\ref{fig:1} (the pole at $\sigma=0$ is removed by the factor $\sigma$).

As we rotate the $\sigma$ ray anticlockwise, past $\arg \sigma=\pi/2$, we pick all the singularities shown in the bottom panel  of Fig.\,\ref{fig:1} giving rise to
\be
\lbl{adlerminkowski}
\mathcal{A}(q^2)=-q^2\int_{\arg\sigma=\pi}\!\!\!\!\!\!d\sigma\
\re^{\sigma q^2} \sigma\borel+\mathcal{A}_{DV}(q^2)\ ,
\quad (q^2>0)\ ,
\ee
where the ``duality violating'' contribution
$\mathcal{A}_{DV}(q^2)$ has been defined as
\be
\lbl{adlerDV}
\mathcal{A}_{DV}(q^2)=\int_{\Gamma} d\sigma\
\re^{\sigma q^2}\sigma\borel\  ,
\ee
and the contour $\Gamma$ is shown in the bottom panel of Fig.\,\ref{fig:1}. Equation (\ref{adlerminkowski}) says that the Adler function defined on the left hand side using  the ray $\arg\sigma=0$ (which is valid for $q^2<0$, cfr. Eq. (\ref{eq:dispersion})) equals the Adler function defined for $\arg\sigma=\pi$ (which is valid for $q^2>0$) \emph{plus} an extra term, given by $\mathcal{A}_{DV}$. As an expansion of $\borel$ in powers of $\sigma$ quickly reveals, both representations of the Adler function produce the \emph{same} OPE in their region of validity in $q^2$, and this says that $\mathcal{A}_{DV}$ is indeed the extra contribution added by the analytic continuation between $q^2 <0$ and $q^2>0$.

Furthermore, using the residue theorem  this extra contribution can be quickly computed to be
\be
\lbl{adlerminkowskitoy}
\mathcal{A}_{DV}(q^2)=2 i\pi
\left(-q^2\frac{d}{dq^2}\right)\sum_{k=1}^{\infty} \re^{i 2 k\pi q^2}
=-\pi \left(-q^2\frac{d}{dq^2}\right)\left(\cot \pi q^2 + i \right)\ ,
\ee
whereby one can see that the DV term encodes the information from the spectrum. Indeed, taking into account the poles of the $\cot \pi q^2$ one finds that \cite{us}
\be
\lbl{checkspectrum}
\frac{1}{\pi}\,\mathrm{Im}\,\Pi(q^2+i\epsilon)
=\sum_{n=1}^{\infty} \delta(q^2-n)\ , \quad (q^2>0) \ ,
\ee
which is the spectrum  in Eq.\,(\ref{massdecay}).

As Fig.\,\ref{fig:1} shows, the singularities in $\borel$ and those of the spectrum are tightly connected. When the $\sigma$ ray reaches $\arg \sigma =\pi/2$ the region covered in $q^2$ touches the positive real axis where the poles in the spectral function are located. If the function $\borel$ did not have these singularities, it would be possible to analytically continue the OPE from $q^2<0$ trivially to $q^2>0$, i.e. on top of the spectrum with all its poles, as if these physical poles did not exist, which clearly cannot be. This means that we can indirectly infer from the position of the spectrum where the singularities of $\borel$ will be located. Furthermore,  what matters is the asymptotic behavior of the spectrum at high excitation number, $n$, since any finite number of terms in the infinite sum (\ref{rhohattoy}) cannot produce a singularity in $\borel$.

Although our model (\ref{rhotoy}) is very illustrative, it is regrettably too simple to reproduce some important features of QCD we cannot simply dismiss. In particular it totally lacks perturbation theory beyond the trivial parton model result $\mathcal{A}=1$. Therefore, we need to generalize it and we will do this in the next section.

\subsection{Exploring the real case at $\mathbf{N_c=\infty}$}
\label{sec:3}

As we have mentioned above, the spectrum in Eq.\,(\ref{massdecay}) has the correct asymptotic behavior expected at high excitation number and at $N_c=\infty$ in QCD, but it is still too simple. In real life one should expect subasymptotic corrections.

Using QCD in two dimensions for guidance \cite{FLZ}, we generalize the spectrum at large $n$ as
\bea
\lbl{reggespectrum}
F(n)&=&1+\epsilon_F(n)\  ,\\
M^2(n)&=&n+ b\log n+ c+ \epsilon_M(n)\ ,\nn
\eea
where $b$ and $c$ are constants, and
\bea
\lbl{corrections}
\epsilon_{i}(n)&=&\epsilon_{i}(0,n)+ \epsilon_{i}(\{\lambda\},n)
\ ,\quad  i=F, M\quad  ,\\
\epsilon_{i}(0,n)&=&\sum_{\nu_{i}>0}\frac{d^{(i)}(\nu_{i})}
{(\log n)^{\nu_{i}}}\ ,
\nonumber\\
\epsilon_{i}(\{\lambda\},n)&=&
\sum_{\lambda_{i}> 0,\, \nu_{i}\gtreqless 0}\frac{d^{(i)}(\lambda_{i},\nu_{i})}
{n^{\lambda_{i}} (\log n)^{\nu_{i}}}\ .
\nonumber
\eea
With the set of values for $\nu_i$ and $\lambda_i$ as chosen in these sums, the
$\epsilon_{i}(n)\rightarrow 0 $ when $n\to \infty$ and they are subleading. It can be seen, and we will discuss it a bit in what follows, that the correction $\epsilon_{i}(0,n)$  will generate the logarithms
that appear in perturbation theory, whereas the
$\epsilon_{i}(\{\lambda\},n)$ corrections with non-zero $\lambda$
contribute to the power corrections \cite{us}.

Clearly, the corrections in Eqs. (\ref{corrections}) are not the most general ones. However, the idea is not to solve the spectrum of QCD by matching the Regge behavior to the OPE but to show whether this could be done \emph{in principle} and, also, whether the main lessons about DVs learned in the previous section remain valid in this case  as well.

In order to do this, as emphasized above, we need to study the behavior of $\borel$ for $\s \to 0^+$. Extensive use of the Converse Mapping Theorem \cite{Flajolet} allows one to express the function $\borel$ given by \footnote{Expressions such as  (\ref{dirichlet}) are known as a Dirichlet series \cite{Dirichlet}. To our knowledge the first paper to connect these series to large-$N_c$ QCD was Ref. \cite{deRafael}.}
\be
 \lbl{dirichlet}
 \borel=\sum_{n=1}^\infty F(n)\, \re^{-\sigma M^2(n)}\ ,
\ee
as \cite{us}
\be
 \lbl{split}
 \borel=\bl+ \bh\ ,
 \ee
where $\bh$ is still to be further split as
\be
\bh=\bh_{\mathrm{Pert. Th.}}+\bh_{\mathrm{OPE-log}}\ .
\ee
As explained in Ref. \cite{us}, both $\bl$ and $\bh_{\mathrm{OPE-log}}$ give rise to terms of the form of the condensate expansion, with the latter giving rise to log corrections to the Wilson coefficients. As to $\bh_{\mathrm{Pert. Th.}}$, it gives rise to the perturbative expansion; it reads

\bea
\lbl{bpt}
\s\bh|_{\mathrm{Pert. Th.}}&=& 1+ \sum_{\nu>0} d^{F}(\nu) \int_0^\infty du
\  \frac{u^{\nu-1}}{\Gamma(\nu)}\ \Gamma(1-u)\,\s^u\
 \nn \\
& \equiv & 1+  \int_0^\infty du\ B^{[\rho]}_{\rm PT}(u)
\  \Gamma(1-u) \, \sigma^{u}\ .
\eea
It is interesting to discuss how $\alpha_s$ is actually hidden in this expression and, with  it, the perturbative expansion. Notice that the integrand contains the function $\Gamma(1-u)$ and one may already suspect that the poles of this function may be related to the renomalon singularities, as it is indeed the case. The presence of the
factor $\sigma^u= \re^{u \log\sigma}$ shows that $\borel$ possesses a cut for $\mathrm{Re}\,\sigma<0$.

Inserting Eq.\,(\ref{bpt}) into Eq.\,(\ref{eq:dispersion}) one obtains
\bea
\lbl{adler2}
\mathcal{A}_{\mathrm{PT}}(q^2)&=&1+ \int_0^\infty du \ B^{[\rho]}_{\rm PT}(u) \left(\frac{\sin \pi u}{\pi u} \right)^{-1} \left( -q^2\right)^{-u}\nn \\
&=&1+ \int_0^\infty du \ B^{[\rho]}_{\rm PT}(u) \left(\frac{\sin \pi u}{\pi u} \right)^{-1}  e^{-u \log \left(-q^2\right)}\ .
\eea
It is important to emphasize that Eq. (\ref{adler2}) contains all the $q^2$ dependence in perturbation theory,\footnote{This expression says that perturbation theory is the part of the large-$q^2$ expansion which depends solely on powers of  $\log(-q^2)$.} irrespective of the definition employed for the coupling $\alpha_s$.

Now it is the time to choose a definition for $\alpha_s$. In Ref. \cite{us} we presented a discussion of perturbation theory in $\alpha_s$ with a beta function limited to just the first coefficient. Now we will generalize this discussion to all orders. In order to do this, we find it convenient to use a definition of this coupling in the  C-scheme of Ref. \cite{C-scheme}.  In the C-scheme the $\beta$ function reads:
\be
\lbl{betaC}
-Q\frac{d\hat{a}_Q}{dQ}=\hat{\beta}=\frac{\beta_1\hat{a}^2_Q}{1-\frac{\beta_2}{\beta_1}\hat{a}_Q}
\ee
where $\hat{a}_Q=\frac{\alpha_s(Q)}{\pi}$. One of the important properties of this C-scheme is that it leads to a rather simple Borel representation of the perturbative series in $\alpha_s$ while being completely general, since any other perturbative definition of $\alpha_s$ can be expressed in terms of the $\alpha_s$ of the C-scheme.

Now, we can trade the dependence on $\log(-q^2)$ for a dependence on  $\hat{a}_Q$ as \cite{C-scheme}
\be
\lbl{logq}
\log \left(-q^2\right)=\frac{1}{\beta_1\hat{a}_Q}+\frac{\beta_2}{\beta_1^2}\log\hat{a}_Q
- \frac{C}{2}
\ee
where the constant $C$ (which is the origin of the name) parametrizes the (residual) scheme dependence within this scheme. Substituting Eq.\,(\ref{logq}) into Eq.\,(\ref{adler2}), one obtains
\be
\lbl{final}
\mathcal{A}_{\mathrm{PT}}(q^2)= 1+ \int_0^\infty du \frac{\mathfrak{B}^{[A]}_{\rm PT}(u)}{\Gamma(1+ \frac{\beta_2}{\beta_1}u)} \ \left( \frac{\hat{a}_Q}{u}  \right)^{-2 u \frac{\beta_2}{\beta_1}}\ e^{-2 u/\beta_1 \hat{a}_Q}
\ee
where
\be
\lbl{def}
\mathfrak{B}^{[A]}_{\rm PT}(u)=\ B^{[\rho]}_{\rm PT}(u) \left(\frac{\sin \pi u}{\pi u} \right)^{-1} \frac{\Gamma(1+ \frac{\beta_2}{\beta_1}u)}{u^{2 \frac{\beta_2}{\beta_1} u}}\ e^{uC}\ .
\ee
The result in Eq. (\ref{final}) reproduces the result in QCD. This (generalized) form for the Borel transform  of the perturbative series in $\alpha_s$  in Eq. (\ref{final}) was first proposed in Ref. \cite{Brown}. Moreover, using the exact relation between $\rho(t)$ and the Adler function:
\be
\rho(t)=\frac{1}{2\pi}\int_0^{2\pi}d\phi\, {\mathcal A}(t e^{i\phi}) \ ,
\ee
in Eq. (\ref{final}), one also obtains an equivalent  Borel representation to (\ref{final}), but now for the spectral function in the C-scheme, with the result
\be
\lbl{def2}
\mathfrak{B}^{[\rho]}_{\rm PT}(u)= \mathfrak{B}^{[A]}_{\rm PT}(u)\ \left(\frac{\sin \pi u}{\pi u} \right)
\ee
 as was also observed in Ref. \cite{Brown}. The conclusion is that our generalized Regge ansatz leads to the expected Borel representation for the perturbative series of QCD as a consequence of the behavior of $\borel$ for $\s \to 0^+$.

 On the other hand, in the previous section we saw how DVs are determined by the singularities of $\borel$ away from the origin. Now, we need to see what happens to those simple poles at $\sigma=\pm  2k\pi i\  (k=1,2,3,\dots)$ when the corrections (\ref{reggespectrum}) are taken into account. Again, use of the Converse-Mapping theorem shows that the main difference is that the simple poles have turned into branching points, leading to \cite{us}
 \be
\lbl{Newcorr2}
\mathcal{A}_{\rm DV}(q^2)= 2\pi i \left(-q^2 \frac{d}{dq^2}\right)\sum_{k=1}^{\infty}
\re^{-4 k \pi^2b}\, \re^{ 2  \pi ik \left(q^2- b \log q^2 -c \right)}
\left(1 + \mathcal{O}\left( \frac{1}{\log q^2}\right)\right)\ .
\ee
This result is still valid for  $N_c=\infty$, although now the Regge spectrum also contains subleading corrections to
exact linear trajectories. These corrections were essential to reproduce the Borel representation of QCD perturbation theory. However, since our world lives at $N_c=3$, we now need to discuss this situation.

\begin{figure}
\begin{center}

\resizebox{0.8\columnwidth}{!}{%
\includegraphics{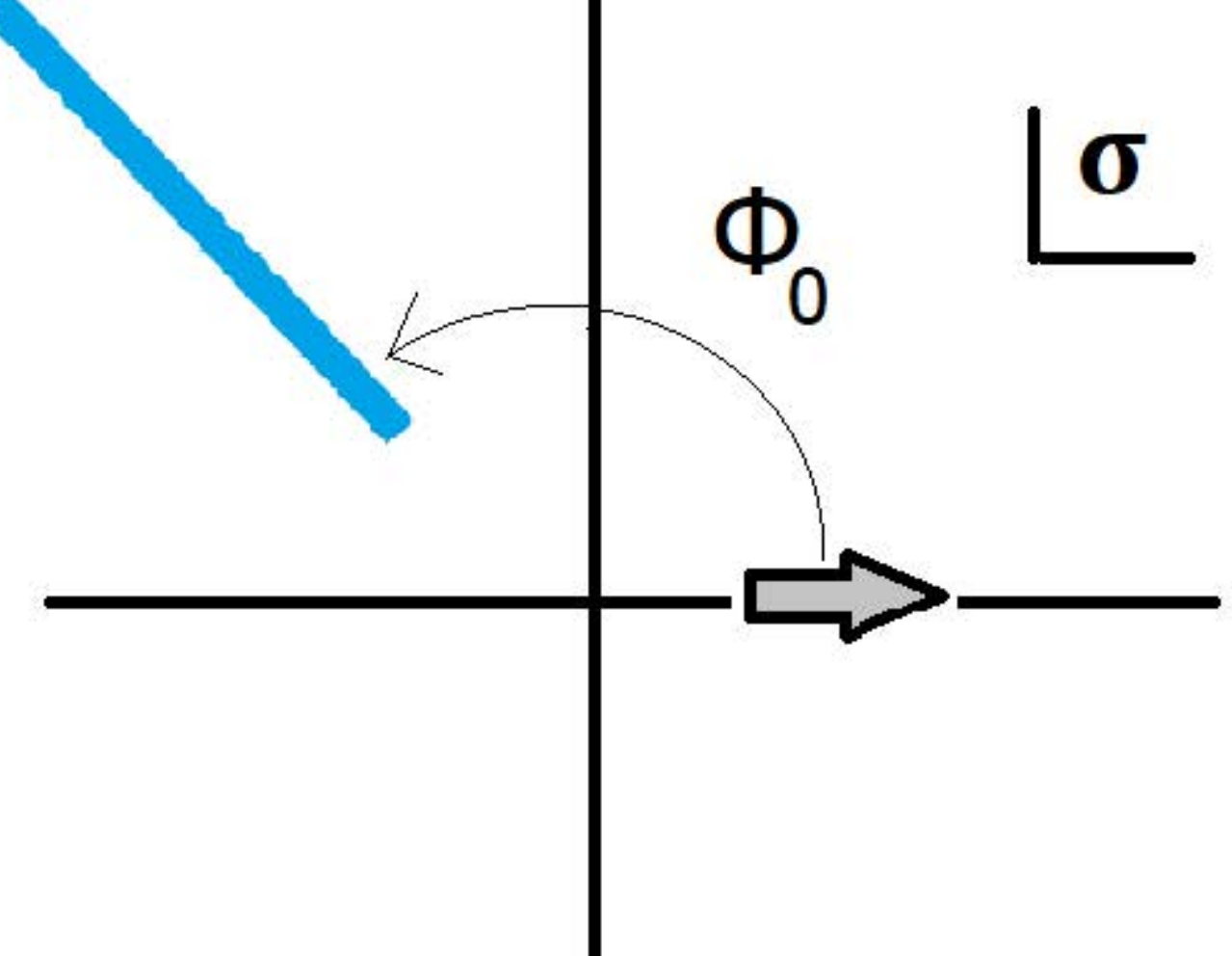}
\hspace{5in}
\includegraphics{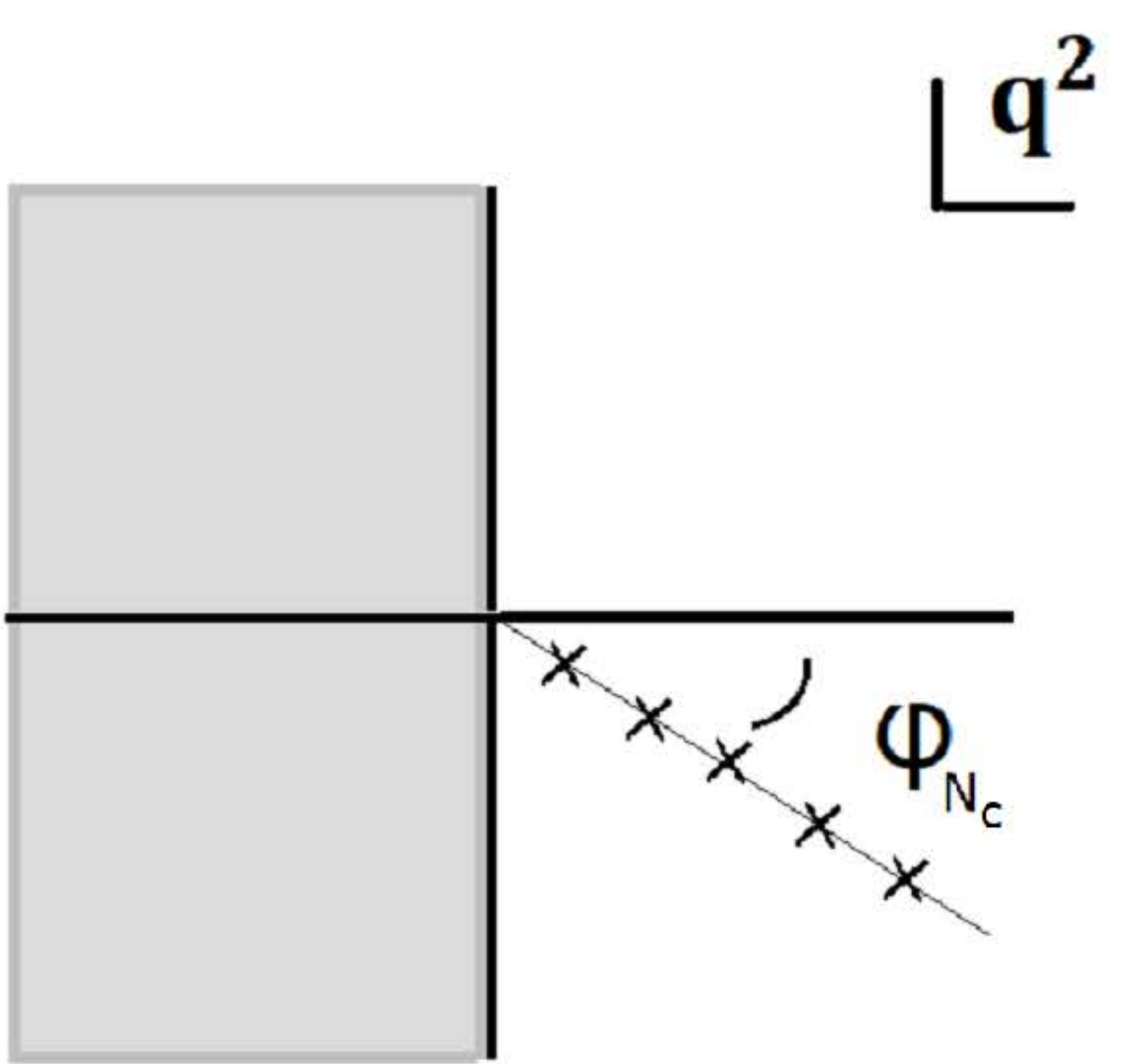}}

\resizebox{0.8\columnwidth}{!}{%
\includegraphics{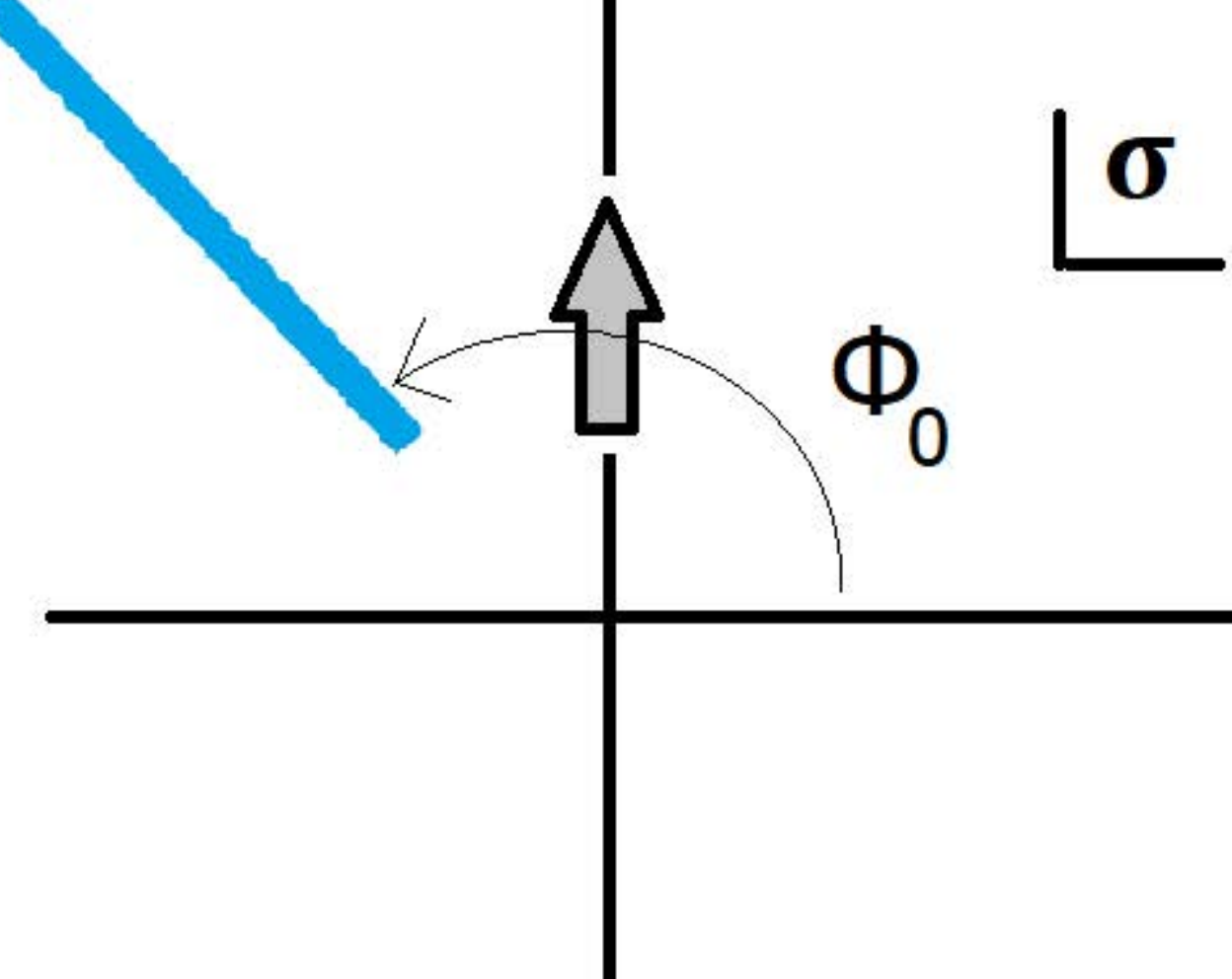}
\hspace{5in}
\includegraphics{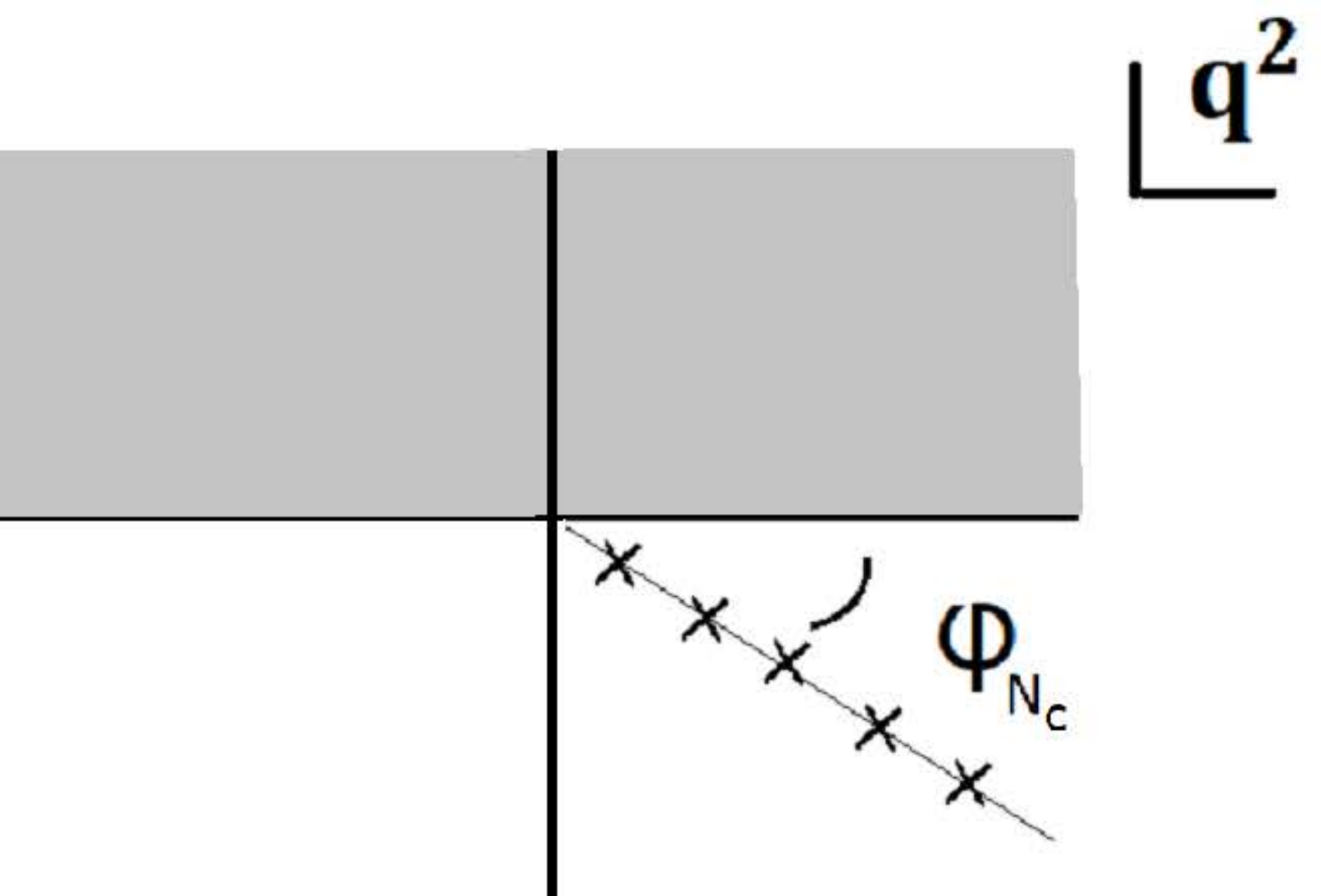}}

\resizebox{0.8\columnwidth}{!}{%
\includegraphics{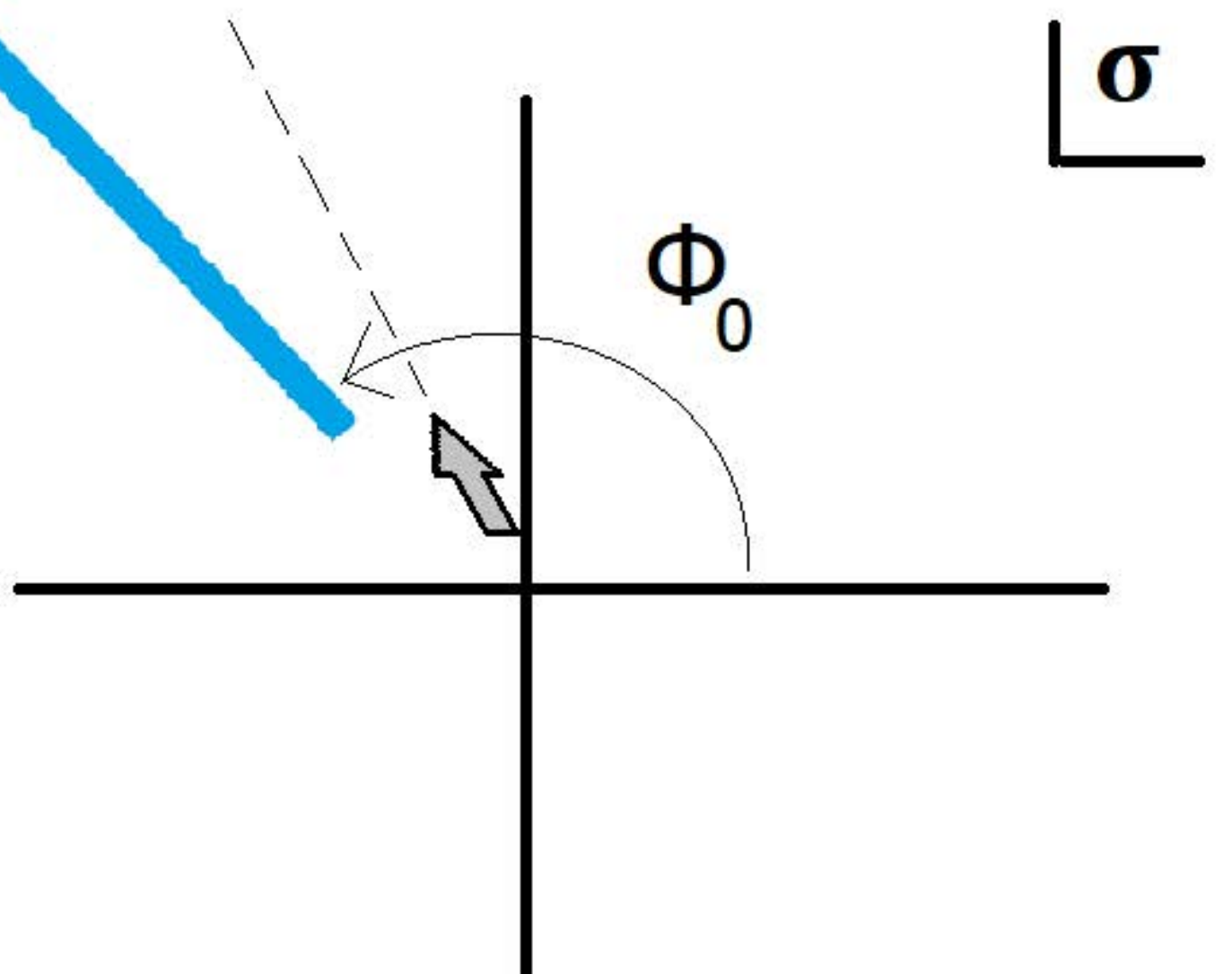}
\hspace{5in}
\includegraphics{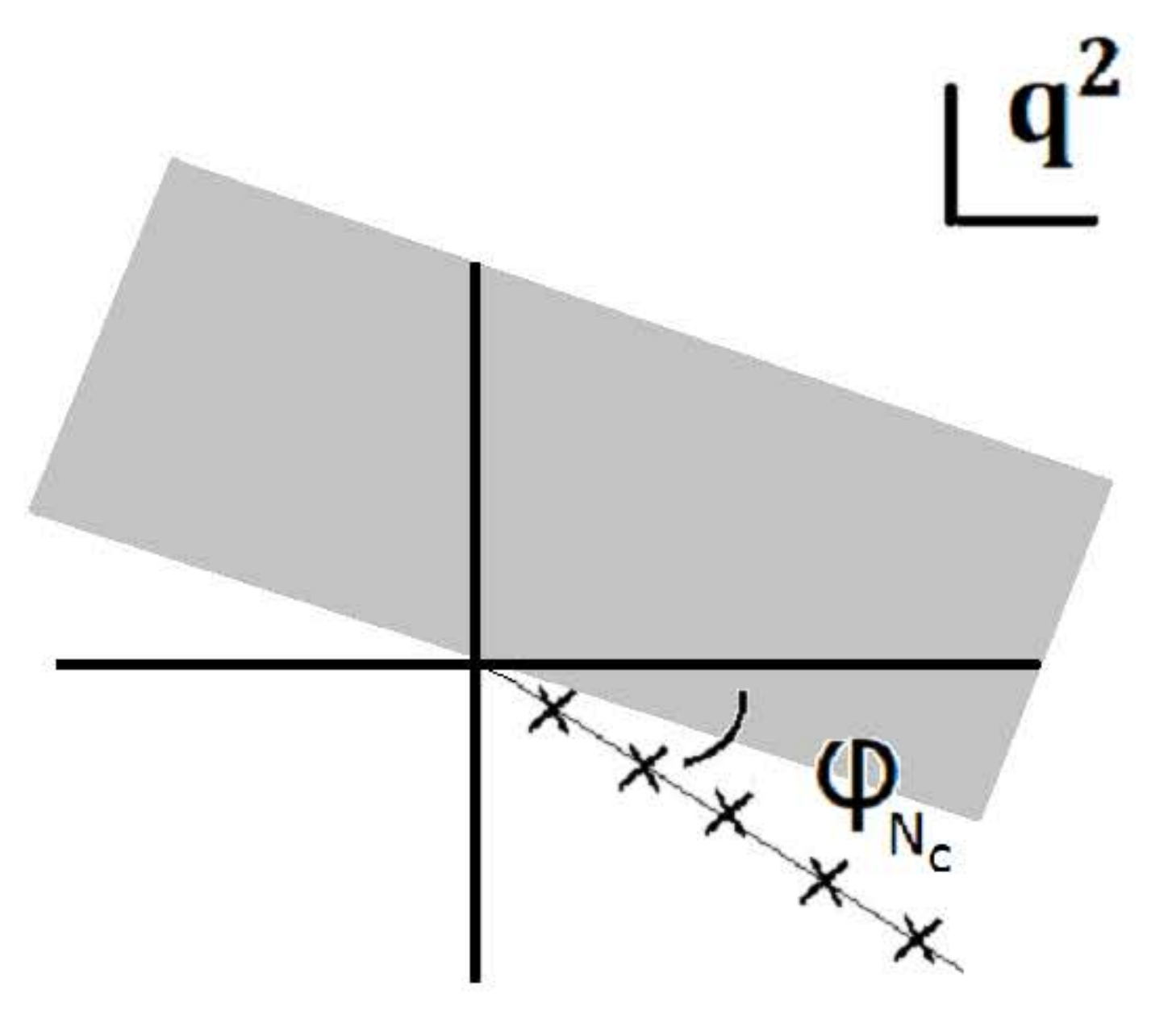}}

\resizebox{0.8\columnwidth}{!}{%
\includegraphics{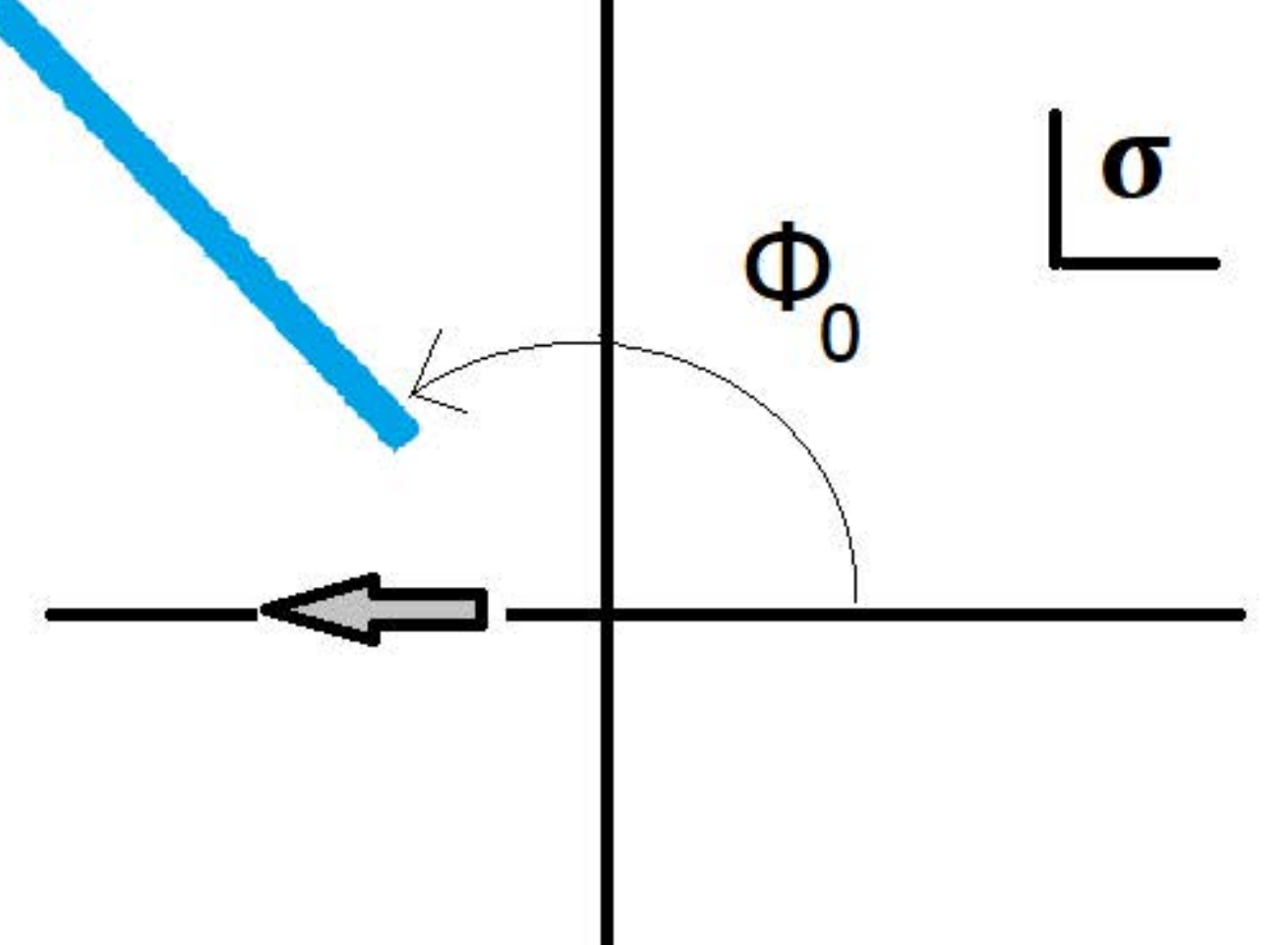}
\hspace{5in}
\includegraphics{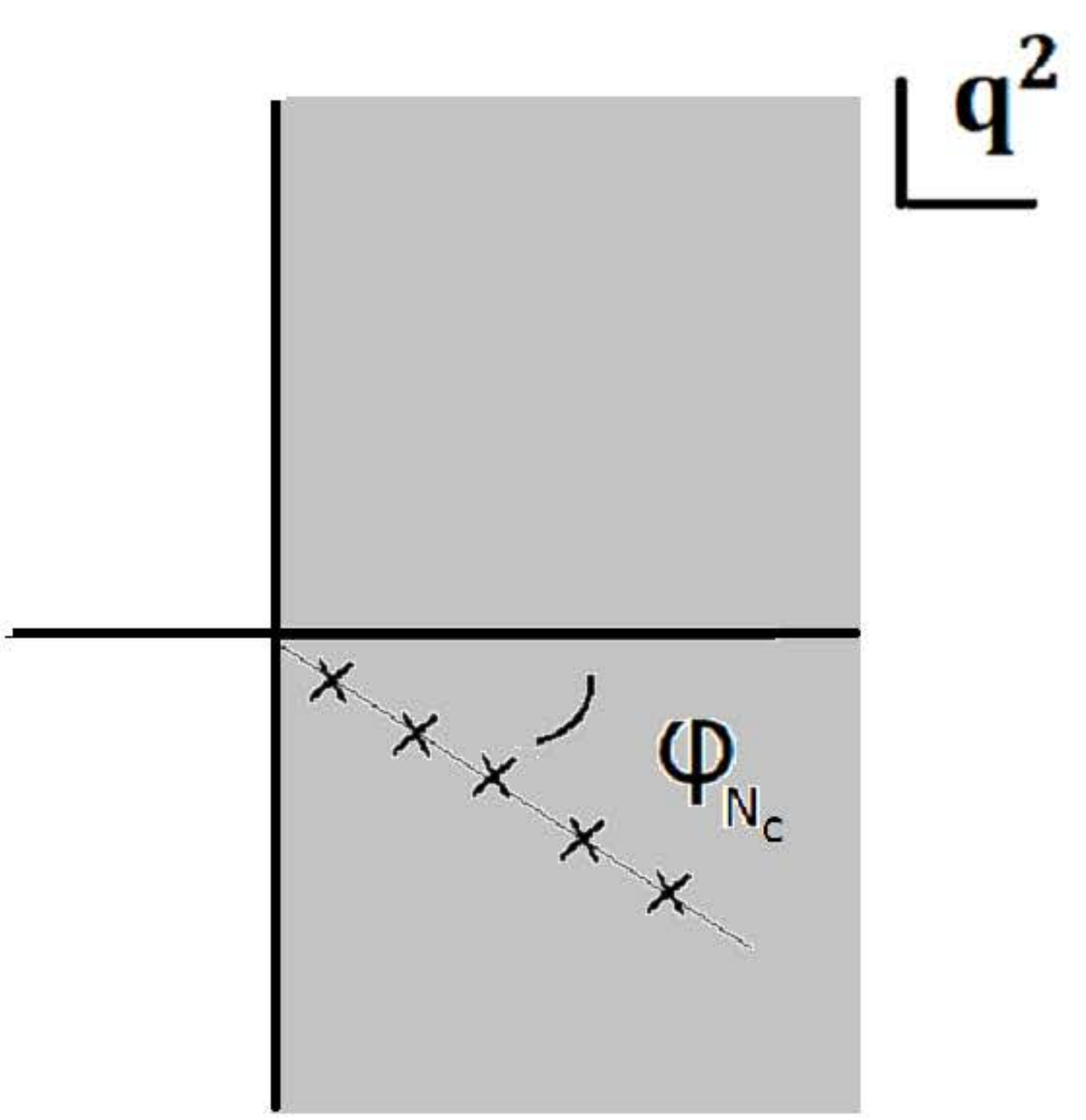}}

\vspace{.5cm}

\hspace{1cm}
\resizebox{0.4\columnwidth}{!}{%
\includegraphics{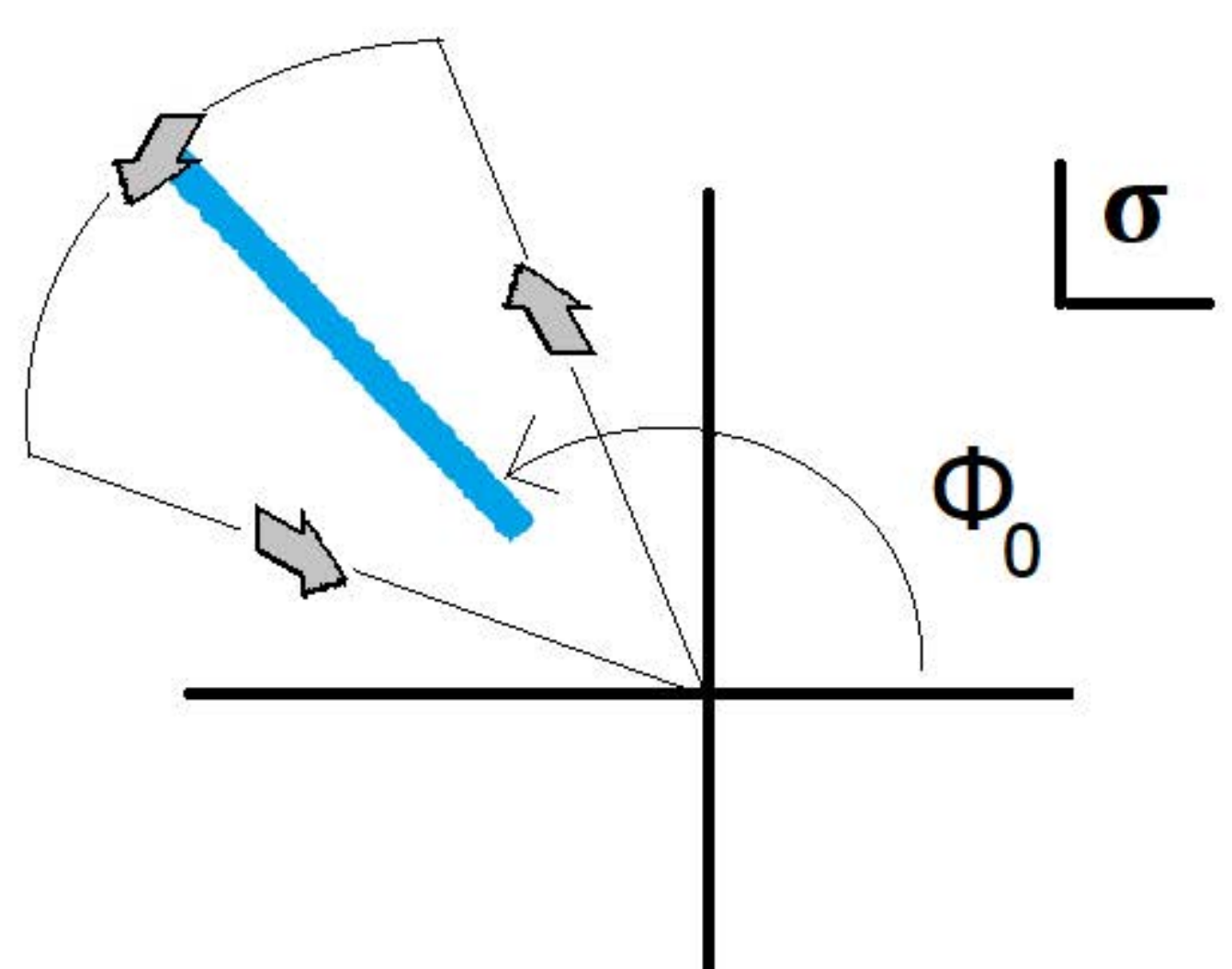}}

\caption{Figure equivalent to Fig. \ref{fig:1} but with finite-$N_c$ corrections included.
The blue line denotes a cut in the $\borel$ function, whose branch point is responsible for the DVs. The angle $\Phi_0\simeq  \frac{\pi}{2}+ \frac{a}{N_c}$. See text. There is another cut starting at the origin along the negative real axis in the $\s$ plane (not shown) which is responsible for perturbation theory.}
\end{center}
\label{fig:2}
\end{figure}

\section{The situation for QCD at $\mathbf{N_c=3}$}
\label{sec:4}

Unlike in the previous sections, our analytic knowledge of the QCD spectrum for $N_c=3$ is not good enough to allow us to calculate the function $\borel$. Therefore, our discussion will necessarily have to be of a qualitative type, based on the fact that $N_c=3$ is close to $N_c=\infty$ \cite{Nc} and our previous analysis will represent a good starting point.

When $N_c$ is considered as large, but finite, a decay width appears and the erstwhile poles on the Minkowski axis recede to the next Riemann sheet below, through the cut in $\mathrm{Im}\Pi(q^2)$ which is created on this axis.

As we rotate the $\sigma$ ray anticlockwise, nothing happens when the $q^2$ region touches the Minkowski axis because the function $\mathcal{A}(q^2)$ is now continuous across the cut into the next Riemann sheet. This is in contradistinction to the $N_c=\infty$ case where those poles were located on this axis causing the $\borel$ function to have singularities on the imaginary axis, as we have discussed.

When $N_c$ is large, but finite, those resonance poles in the complex $q^2$ plane recede by an angle approximately given by
\be
\lbl{angle}
\tan \varphi_{N_c}\approx \varphi_{N_c}  =- \frac{\Gamma}{M}= -
\frac{a}{N_c}\left(1 + \mathcal{O}\left(\frac{1}{N_c}\right)\right)\ ,
\ee
where $a\sim N_c^0>0$, $\Gamma\sim 1/N_c$ and $M\sim N_c^0$. Both the string picture \cite{string}, as well as two-dimensional QCD \cite{Blok}, support that the parameter $a$ is independent of the resonance excitation number $n$. Then, according to Eq.~(\ref{eq:dispersion}), as these finite-$N_c$ corrections push the resonance poles an angle
$\varphi_{N_c}\approx - \frac{a}{N_c}$ into the next lower Riemann sheet in the complex $q^2$ plane,
the singularities of the $\borel$ function rotate anticlockwise past the positive imaginary axis in the $\s$ plane and position themselves   (approximately) along the ray $\arg \sigma \simeq  \frac{\pi}{2}+ \frac{a}{N_c}\equiv\Phi_0$
creating a series of branch points (see Fig.~2). Each of these cuts starts at a  branch point and it is the closest of these points to the origin, $\sh$, which controls the exponent in the exponential falloff of the DVs. From our previous discussion of the situation at $N_c=\infty$ we know that this point is at a distance $|\sh|=2\pi (1+ \mathcal{O}(N_c^{-1}))$ so we can estimate that the closest branching point to the origin is located at $\sh=2\pi \mathrm{e}^{i \Phi_0}  (1+ \mathcal{O}(N_c^{-1}))$ with $\Phi_0=\frac{a}{N_c}+ \frac{\pi}{2}$ (see Fig. 2).

Again, DVs can be obtained by calculating the contribution from the contour depicted on the bottom panel in Fig. 2. The results gives \cite{us}
\bea
\lbl{resultNc}
\mathcal{A}_{\rm DV}(q^2)&=&  2 \pi  i \left(-q^2 \frac{d}{dq^2}\right) \re^{-2\pi^2 b}\, \re^{-2\pi ci}\,
(-q^2)^{-2\pi ib}\ \re^{ 2\pi q^2 \left(i-\frac{a}{N_c}\right)}\ \nn\\
&&\hspace{-1cm}\times  \left[1+\mathcal{O}\left(\frac{1}{q^2}\right)
+ \mathcal{O}\left(\frac{1}{\log q^2}\right)  \right]+
\mathcal{O}\left(\frac{1}{N_c}\right)\ ,
\eea
from which one can extract the leading contribution to the DV part of the spectral function as\footnote{Recall that our momentum is in units of $\Lambda_{QCD}$.}
\bea
\lbl{final result}
 \hspace{-3cm}\mathrm{Im}\Pi_{\rm DV}(q^2)&\approx & \ 2\pi \,  \re^{-4\pi^2 b}\,
\re^{-2 \pi q^2 \frac{a}{N_c}} \cos2\pi\Big(q^2- c- b \log{q^2}\Big)\\
  &&  \times
   \left(1+ \mathcal{O}\left(\frac{1}{N_c};
\frac{1}{q^2};\frac{1}{\log q^2}\right)\right)\nn\ .
 \eea

This result for $\mathrm{Im}\Pi_{\rm DV}$ is to be considered the leading correction  to the condensate expansion from the OPE on the Minkowski axis, at large $q^2$ right below the perturbative regime, and shows how the theoretical spectral function is corrected away from perturbation theory (even if enhanced with the condensate expansion).

\section{Conclusions}
\label{sec:5}

We would like to emphasize a few important lessons which may be extracted from our analysis. The properties of the OPE as an expansion are encoded in the function $\borel$ which has a very rich singularity structure in the $\s$ complex plane. Apart from the cut  starting at the origin of the negative real axis, which controls all the logarithmic corrections to the OPE (including the whole perturbative series in $\alpha_s$), there will also be branch point singularities, $\sh$,  in the left half plane located at a finite distance from the origin, but close by an amount $\sim 1/N_c$ to the imaginary axis.

If, as it is sometimes done, one is willing to consider the pure power corrections of the OPE as a standalone series, because of $|\sh| \neq 0$,  the sub-series of $\borel$ in powers of $\s$ will have a finite radius of convergence and, consequently, these power corrections for the Adler function in Eq. (\ref{eq:dispersion}) will miss terms of order $\exp(-|\sh| q^2)$, even in the Euclidean. However, DVs originate not from the $|\sh|$ but from $\mathrm{Re}\,\sh$ which, as we have discussed, is negative, yielding
$\mathrm{Im}\Pi_{DV} \sim \exp{( \mathrm{Re}\,\sh)} \sim \exp{( -\Gamma/M)} $ where $\Gamma/M\sim a/N_c$ is the angle receded by the resonances very high up in the spectrum, as given in  Eq. (\ref{angle}).

In a series of papers \cite{Cata1}-\cite{Boito4} the determination of $\alpha_s$ from tau decays was reanalyzed with the use of an expression like (\ref{final result}), but without the $\log q^2$ correction in the argument of the oscillation. We remark that a $\log q^2$ has a very slow variation in a finite interval in momentum in comparison with the $q^2$. This means that this $\log q^2$ effectively amounts to a renormalization of the constant $c$ in Eq. (\ref{final result}). Furthermore, all phenomenological analysis of the radial Regge behavior of the QCD spectrum has not found any evidence for a non-zero $b$ term.

To be more precise, a fit of the radial Regge trajectories to the meson spectrum finds the values \cite{Masjuan}
\be
\Lambda^2 \simeq 1.35(4)\, \mathrm{GeV}^2\quad , \quad \frac{\Gamma}{M}\simeq 0.12(8) \left( \simeq \frac{a}{N_c}\right)\ .
\ee
Reintroducing the $\Lambda$ scale into the expression (\ref{final result}) and parametrizing the DV component of the spectral function as \cite{Cata1}
\be
\rho_{\mathrm{DV}}(t)\propto e^{-\gamma t} \sin\left(\alpha+\beta t\right)
\ee
we obtain
\bea
\lbl{number1}
\beta&=&\frac{2 \pi}{\Lambda^2}\left[1 + \mathcal{O}\left(\frac{1}{N_c}\right)\right]\simeq 4.7(2)\, \mathrm{GeV}^{-2} \left[1+ \mathcal{O}\left(\frac{1}{N_c}\right)\right]\nn \\
\gamma&=&\frac{2 \pi}{\Lambda^2}\frac{a}{N_c} \left[1+ \mathcal{O}\left(\frac{1}{N_c}\right)\right]\simeq 0.6(4)\left[1 +\mathcal{O}\left(\frac{1}{N_c}\right)\right]\ .
\eea
These may be compared to the results from a recent fit to an improved  $\tau$ vector isovector spectral function, obtained from a combination of ALEPH and OPAL two- and four-pion exclusive modes, BaBar $\tau$-decay results for the $K^-K^0$ mode and CVC evaluations of $e^+e^-\to hadrons$ cross section for the remaining exclusive modes \cite{Boito4}:
\be
\lbl{number2}
\beta=3.81(26)\, \mathrm{GeV}^{-2} \quad ,\quad \gamma=0.57(17)\ \mathrm{GeV}^{-2} \ .
\ee
The agreement between Eq.\,(\ref{number1}) and Eq.\,(\ref{number2}) is reassuring. Furthermore, the $2\pi$ factors in (\ref{number1}) reflect the fact that this agreement cannot be simply the result of dimensional analysis, and gives further theoretical support to the parametrization of DVs  employed in  \cite{Cata1}-\cite{Boito4}.

\vspace{2cm}

\textbf{Acknowledgements}

I am very grateful to D.~Boito, I.~Caprini, M.~Golterman and K.~Maltman for innumerable discussions on DVs and a pleasant collaboration. Work supported by CICYTFEDER-FPA2017-86989-P and by Grant No. 2017 SGR 1069.
IFAE is partially funded by the CERCA program of the Generalitat de Catalunya.


\begin{thebibliography}{}

\bibitem{PPR}
See, e.g.,
S.~Peris, M.~Perrottet and E.~de Rafael,
JHEP \textbf{05}, 011 (1998)
doi:10.1088/1126-6708/1998/05/011
[arXiv:hep-ph/9805442 [hep-ph]].




\bibitem{Shif}
  M.~A.~Shifman,
  Int.\ J.\ Mod.\ Phys.\ A 11, 3195 (1996)
  [hep-ph/9511469].

\bibitem{Blok}
  B.~Blok, M.~A.~Shifman and D.~X.~Zhang,
  Phys.\ Rev.\ D 57, 2691 (1998)
  Erratum: [Phys.\ Rev.\ D 59, 019901 (1999)]
  doi:10.1103/PhysRevD.57.2691, 10.1103/PhysRevD.59.019901
  [hep-ph/9709333].


\bibitem{Blok}
  B.~Blok, M.~A.~Shifman and D.~X.~Zhang,
  Phys.\ Rev.\ D 57, 2691 (1998)
  Erratum: [Phys.\ Rev.\ D 59, 019901 (1999)]
  doi:10.1103/PhysRevD.57.2691, 10.1103/PhysRevD.59.019901
  [hep-ph/9709333].




\bibitem{Shif1}
  M.A. Shifman,
  in {\it At the frontier of particle physics},
vol. 3, 1447-1494, World Scientific, 2001 [arXiv:hep-ph/0009131].





\bibitem{PQW}
E.~C.~Poggio, H.~R.~Quinn and S.~Weinberg,
Phys. Rev. D \textbf{13}, 1958 (1976)
doi:10.1103/PhysRevD.13.1958



\bibitem{Shifman}
M.~Shifman,
eConf \textbf{C030614}, 001 (2003)
WQCD-2003-001,
and references therein.


\bibitem{us}
D.~Boito, I.~Caprini, M.~Golterman, K.~Maltman and S.~Peris,
Phys. Rev. D \textbf{97}, no.5, 054007 (2018)
doi:10.1103/PhysRevD.97.054007
[arXiv:1711.10316 [hep-ph]].



\bibitem{math} See, for example,
I.~Aniceto, G.~Basar and R.~Schiappa,
Phys. Rept. \textbf{809}, 1-135 (2019)
doi:10.1016/j.physrep.2019.02.003
[arXiv:1802.10441 [hep-th]];
T.M. Seara and D. Sauzin,
\emph{Resumaci\'{o} de Borel i Teoria de la Ressurgencia}
(in Catalan), Butlleti de la Societat Catalana de Matem\`{a}tiques
\textbf{18} (2003) 129; B. Candelpergher et al., \emph{Approche de
la r\'{e}surgence}, Actualit\'{e}s Math\'{e}matiques, Hermann edit.
(1993), in particular section ``Commentaire 1: Sommation de Borel.''
For a rather exhaustive treatise, see O. Costin, \emph{Asymptotics and
Borel Summability}, Chapman and Hall/CRC, Monographs and Surveys in
Pure and Applied Mathematics, in particular, section 4.4d.

\bibitem{tHooft2d} G.~'t Hooft,
Nucl.\ Phys.\ B 75, 461 (1974).

\bibitem{CCG} C.~G.~Callan, Jr., N.~Coote and D.~J.~Gross,
Phys.\ Rev.\ D 13, 1649 (1976).

\bibitem{string} See, for example,  M.~Shifman and A.~Vainshtein,
  Phys.\ Rev.\ D 77, 034002 (2008)
  [arXiv:0710.0863 [hep-ph]],
  and references therein.



\bibitem{Masjuan} See, for example, P.~Masjuan, E.~Ruiz Arriola and
W.~Broniowski,
Phys.\ Rev.\ D 85, 094006 (2012)
[arXiv:1203.4782 [hep-ph]].

\bibitem{FLZ} V.~A.~Fateev, S.~L.~Lukyanov and A.~B.~Zamolodchikov,
J.\ Phys.\ A 42, 304012 (2009)
[arXiv:0905.2280 [hep-th]].


\bibitem{Flajolet} P. Flajolet {\it et al.}, Theoretical Computer Science
144 (1995) 3.



\bibitem{Dirichlet} See, for example, G.H. Hardy and M. Riesz,
{\it The General Theory of Dirichlet's Series}, Cambridge Tracts
in Mathematics and Mathematical Physics, Cambridge Univ. Press (1915).
A useful summary can be found in
https://en.wikipedia.org/wiki/General-Dirichlet-series,
and in https://en.wikipedia.org/wiki/Dirichlet-series.



\bibitem{deRafael} E.~de Rafael,
Nucl.\ Phys.\ Proc.\ Suppl.\  207-208, 290 (2010)
[arXiv:1010.4657 [hep-th]].

\bibitem{C-scheme}
  D.~Boito, M.~Jamin and R.~Miravitllas,
  Phys.\ Rev.\ Lett.\  {\bf 117}, no. 15, 152001 (2016)
  doi:10.1103/PhysRevLett.117.152001
  [arXiv:1606.06175 [hep-ph]].


\bibitem{Brown}
 L.~S.~Brown, L.~G.~Yaffe and C.~X.~Zhai,
  Phys.\ Rev.\ D {\bf 46}, 4712 (1992)
  doi:10.1103/PhysRevD.46.4712
  [hep-ph/9205213].




\bibitem{Nc}
         G~'t Hooft, Nucl. Phys. {\bf B72} (1974) 461; {\bf B73} (1974) 461;
         E.~Witten, Nucl. Phys. {\bf B79} (1979) 57.



\bibitem{Cata1}
  O.~Cat\`a, M.~Golterman and S.~Peris,
  JHEP 0508, 076 (2005)
  [hep-ph/0506004].
 	
\bibitem{Cata2}
  O.~Cat\`a, M.~Golterman and S.~Peris,
  Phys.\ Rev.\ D 77, 093006 (2008)
  [arXiv:0803.0246 [hep-ph]].

\bibitem{Boito1}
  D.~Boito, O.~Cat\`a, M.~Golterman, M.~Jamin, K.~Maltman, J.~Osborne and S.~Peris,
  Phys.\ Rev.\ D 84, 113006 (2011)
  [arXiv:1110.1127 [hep-ph]].

\bibitem{Boito2}
  D.~Boito, M.~Golterman, M.~Jamin, A.~Mahdavi, K.~Maltman, J.~Osborne and S.~Peris,
  Phys.\ Rev.\ D 85, 093015 (2012)
  [arXiv:1203.3146 [hep-ph]].

\bibitem{Boito3} D.~Boito, M.~Golterman, K.~Maltman, J.~Osborne and S.~Peris,
Phys.\ Rev.\ D 91, 034003 (2015)
[arXiv:1410.3528 [hep-ph]].




\bibitem{Boito4}
D.~Boito, M.~Golterman, K.~Maltman, S.~Peris, M.~V.~Rodrigues and W.~Schaaf,
Phys. Rev. D \textbf{103}, no.3, 034028 (2021)
doi:10.1103/PhysRevD.103.034028
[arXiv:2012.10440 [hep-ph]].



\end{thebibliography}
\end{document}